\documentclass[mnsc,nonblindrev]{informs3-YCA}

\OneAndAHalfSpacedXI 

\usepackage{endnotes}
\let\footnote=\endnote

\usepackage{url}
\usepackage{enumitem}
\usepackage{amsmath}
\usepackage{amssymb}
\usepackage{natbib}
\bibpunct[, ]{(}{)}{,}{a}{}{,}%
\def\bibfont{\small}%
\def\bibsep{\smallskipamount}%
\usepackage{tikz}
\usepackage{pgfplots}
\usetikzlibrary{arrows}
\usetikzlibrary{calc,arrows.meta,positioning}
\usepackage{ctable}
\usepackage{booktabs}
\usepackage{multirow}
\usepackage{algorithm}
\usepackage[noend]{algorithmic}
\usepackage{color}
\usepackage{graphicx}
\usepackage{enumitem}
\usepackage{cancel}

\allowdisplaybreaks

\usepackage{mathtools}
\usepackage[hidelinks]{hyperref}

\usepackage{xspace}

\usepackage{pgfplotstable}
\usetikzlibrary{arrows.meta}
\usepgfplotslibrary{groupplots}

\usepackage[title]{appendix}
\usepackage{titletoc}
\usepackage{accents}

\newcommand{\expar}[1]{{\mathbb E} [ #1 ]}

\newcommand{\rev}{{\cal R}}

\usepackage{titlesec}
\titleformat{\paragraph}[runin]{\normalfont\bfseries}{\theparagraph}{1em}{}

\usepackage{natbib}
 \bibpunct[, ]{(}{)}{,}{a}{}{,}%
 \def\bibfont{\small}%
 \def\bibsep{\smallskipamount}%
\TheoremsNumberedThrough     
\ECRepeatTheorems

\EquationsNumberedThrough    

\newcounter{jakecounter}

\begin{document}


\RUNAUTHOR{Feldman,Huang,Zhang}

\RUNTITLE{Inventory and Assortment Optimization under General Choice}

\TITLE{\large Inventory and Assortment Optimization under General Choice:
A Unified Approximation Framework}

\ARTICLEAUTHORS{%
\AUTHOR{Jacob Feldman}
\AFF{Washington University in St. Louis}
\AUTHOR{Yukai Huang}
\AFF{Suffolk University}
\AUTHOR{Heng Zhang}
\AFF{Arizona State University}
} 

\ABSTRACT{We study joint inventory and assortment optimization under general
substitutable choice models. Before a selling horizon, a decision maker
selects initial product inventories subject to a capacity constraint
on the total number of units stocked. Customers subsequently arrive
sequentially and choose among the products offered to them. We consider
two operating settings: show-all inventory selection, in which every
in-stock product must be offered, and dynamic assortment personalization,
in which the decision maker observes each customer's type and may
adapt the offered assortment to the remaining inventory and purchase
history. 

We develop a unified approximation framework that connects each
operating setting to its corresponding static optimization problem.
For show-all inventory selection, a polynomial-time
$\alpha$-approximation for cardinality-constrained static assortment
optimization yields guarantees of $0.486\alpha$ when the number of
arrivals is deterministic or Poisson distributed and $0.4328\alpha$
when it follows an increasing-failure-rate (IFR) distribution.
For personalization, the same guarantees hold when $\alpha$ instead
denotes the approximation factor for joint assortment and customization,
the static problem studied by El Housni and Topaloglu: selecting a
common product catalog subject to a cardinality constraint and a
customized assortment from that catalog for each customer type.
Our approach combines a fluid relaxation, demand-dependent fractional
inventory allocations, bounds on expected truncated demand, and
dependent rounding. Notably, our personalized policy fixes one
assortment for each customer type before the selling horizon and
subsequently offers every in-stock product within that assortment.
Thus, it requires no dynamic assortment reoptimization, despite
competing against the optimal fully adaptive policy. 

Finally, we construct instances in which optimal show-all revenue approaches only one-half of the fluid bound under deterministic arrivals and $1-w\approx0.4328567$ under IFR arrivals, where
$we^w=1$. With an exact static assortment oracle, our IFR guarantee
can approach the latter factor arbitrarily 
}%

\maketitle

\vspace{-1.0cm}

\section{Introduction}
\label{sec:introduction}

A fundamental problem in retail operations is to determine which products to
carry and how much inventory to stock before a selling season. These decisions
are particularly challenging when customers substitute following stockouts:
the demand for a product depends on the availability of other products, and
hence on their initial inventory levels and realized sales. We consider two
versions of this problem, both subject to a capacity constraint on the total
number of units initially stocked. In the \emph{show-all} (SA) setting, every
in-stock product is offered to each arriving customer, so the assortment
evolves only through stockouts. In the \emph{dynamic assortment personalization}
(DAP) setting, the decision maker observes each customer's type and can offer
a customized subset of the in-stock products. The latter provides additional
flexibility but requires coordinating a common initial inventory across
customers with heterogeneous preferences.

We develop a unified approximation framework for both problems under general substitutable choice models. Our approach separates the computational difficulty of selecting products from the stochastic difficulty of allocating inventory and managing stockouts. For SA, the relevant static optimization problem is cardinality-constrained assortment optimization. For DAP, it is joint assortment and customization (JAC): selecting a common catalog of at most $B$ products and an assortment from that catalog for each customer type. An approximation algorithm for the appropriate static problem serves as an oracle for a fluid relaxation, whose solution is converted into a feasible integral inventory decision through demand-dependent fractional allocations and randomized rounding. Importantly, our DAP policy fixes an assortment for each customer type before the selling horizon and subsequently offers every in-stock product from that assortment whenever the corresponding type arrives. Thus, although the benchmark permits fully adaptive assortment decisions, our policy requires no dynamic assortment reoptimization: assortments change only as their constituent products stock out. This illustrates how appropriately coordinated initial inventory and assortment decisions can substantially reduce the need for dynamic assortment control.

\paragraph{Related work on show-all inventory selection.}
Early approximation algorithms exploit structure in customer preferences.
\citet{goyal2016near} develop approximation schemes for structured preference
lists, and \citet{segev2019nested} gives a quasi-polynomial-time approximation
scheme for nested preferences. For general rank-based choice,
\citet{aouad2019approx} obtain revenue-dependent guarantees and a constant
factor under uniform product revenues. \citet{goldsteinfeder2023dynamic}
improve this horizontally differentiated case and develop a universal inventory
construction across arrival-count distributions. Under multinomial-logit (MNL)
choice, \citet{aouad2018greedy} establish a constant-factor guarantee for
increasing-failure-rate (IFR) arrivals, and \citet{aouad2023stability}
develop an approximation scheme whose running time depends on the range of
preference weights. More recently, \citet{sun2025unified} provide SA guarantees
of $0.474-\varepsilon$ for deterministic arrivals and $0.286-\varepsilon$ for
IFR arrivals using MNL-specific structure.

Moving beyond specific parametric choice models, \citet{fu2026joint} introduce
a black-box reduction from cardinality-constrained assortment optimization to
show-all inventory selection. Given an $\alpha_{\mathrm{SA}}$-approximation
for the static assortment problem, their guarantees are approximately
$0.272\alpha_{\mathrm{SA}}$ for deterministic or Poisson arrivals and
$0.25\alpha_{\mathrm{SA}}$ for IFR arrivals. We retain their general
substitutability assumption while improving these guarantees to
$0.486\alpha_{\mathrm{SA}}$ and $0.4328\alpha_{\mathrm{SA}}$,
respectively. Our analysis also gives matching or nearly matching examples for
the gap between the stochastic optimum and the fluid relaxation: a factor of
$1/2$ for deterministic arrivals and the sharp factor
$1-w\approx0.4328567$ for IFR arrivals, where $we^w=1$.

Table~\ref{tab:sa_and_oracles} summarizes the SA comparisons alongside the
static assortment oracles used to instantiate our guarantees. For MNL,
cardinality-constrained assortment optimization is polynomially solvable, so
$\alpha_{\mathrm{SA}}=1$ \citep{rusmevichientong2010dynamic}.
For rank-based choice with uniform product revenues, purchase probability is a
weighted coverage function, and greedy optimization gives
$\alpha_{\mathrm{SA}}=1-1/e$ \citep{aouad2019approx}. The latter guarantee
continues to apply to the common price-adjusted rewards in our fluid
separation problem. Our factors therefore specialize to $0.486$ and $0.4328$
for MNL and to approximately $0.3072$ and $0.2736$ for uniform-revenue
rank-based choice, under deterministic/Poisson and IFR arrivals, respectively.

\begin{table}[t]
\centering
\caption{Show-all approximation guarantees and static assortment oracles.}
\label{tab:sa_and_oracles}
\begingroup
\footnotesize
\setlength{\tabcolsep}{3pt}
\renewcommand{\arraystretch}{1.15}
\begin{minipage}[t]{0.65\textwidth}
\vspace{0pt}
\centering
\textbf{(a) Show-all inventory selection}\par\smallskip
\begin{tabular}{@{}
>{\raggedright\arraybackslash}m{0.40\linewidth}
>{\centering\arraybackslash}m{0.27\linewidth}
>{\centering\arraybackslash}m{0.27\linewidth}@{}}
\toprule
\textbf{Paper} & \textbf{Deterministic} & \textbf{IFR} \\
\midrule
\multicolumn{3}{@{}l}{\textit{MNL choice}} \\
\addlinespace[2pt]
\citet{aouad2018greedy} & $0.139^{\dagger}$ & $0.139^{\dagger}$ \\
\addlinespace[3pt]
\citet{sun2025unified} (DP) & $0.474$ & $0.286$ \\
\addlinespace[3pt]
\textbf{This paper} & $\mathbf{0.486}$ & $\mathbf{0.4328}$ \\
\addlinespace[7pt]
\multicolumn{3}{@{}l}{\textit{Rank-based choice; uniform revenues}} \\
\addlinespace[2pt]
\citet{aouad2019approx} & $\approx0.1937$ & $\approx0.1937$ \\
\addlinespace[3pt]
\citet{goldsteinfeder2023dynamic} & $0.226$ & $0.226$ \\
\addlinespace[3pt]
\textbf{This paper} & $\boldsymbol{\approx0.3072}$ & $\boldsymbol{\approx0.2736}$ \\
\addlinespace[7pt]
\multicolumn{3}{@{}l}{\textit{General substitutable choice}} \\
\addlinespace[2pt]
\citet{fu2026joint} & $0.272\alpha_{\mathrm{SA}}$ & $0.25\alpha_{\mathrm{SA}}$ \\
\addlinespace[3pt]
\textbf{This paper} & $\boldsymbol{0.486\alpha_{\mathrm{SA}}}$ & $\boldsymbol{0.4328\alpha_{\mathrm{SA}}}$ \\
\bottomrule
\end{tabular}
\end{minipage}
\hfill
\begin{minipage}[t]{0.32\textwidth}
\vspace{0pt}
\centering
\textbf{(b) Static assortment oracle}\par\smallskip
\begin{tabular}{@{}
>{\raggedright\arraybackslash}m{0.65\linewidth}
>{\centering\arraybackslash}m{0.27\linewidth}@{}}
\toprule
\textbf{Choice model} & $\boldsymbol{\alpha_{\mathrm{SA}}}$ \\
\midrule
MNL & $1$ \\
\addlinespace[3pt]
Rank-based; uniform revenues & $1-1/e$ \\
\bottomrule
\end{tabular}
\par\smallskip
\raggedright
The MNL oracle is exact \citep{rusmevichientong2010dynamic};
the uniform-revenue rank-based problem is weighted maximum coverage
\citep{aouad2019approx}.
\end{minipage}
\par\medskip
\begin{minipage}{\textwidth}
\footnotesize
\textit{Notes.} Arbitrarily small accuracy losses are suppressed, and
reported decimals are rounded. The Aouad et al.\ (2019) factor is
$(e-1)/(4e-2)$. Guarantees in the deterministic column for
\citet{fu2026joint} and this paper also apply to Poisson arrivals.
$\dagger$ The $0.139$ guarantee uses the auxiliary oracle of
\citet{aouad2018greedy}. Assumptions and computational requirements differ
across papers.
\end{minipage}
\endgroup
\end{table}

\paragraph{Related work on assortment personalization.}
For DAP, \citet{bai2025coordinated} study coordinated inventory stocking and assortment customization, providing constant-factor guarantees under MNL choice. \citet{sun2025unified} subsequently improve these guarantees to $\frac12(1-1/e)-\varepsilon$ for deterministic arrivals and $1/4-\varepsilon$ for arbitrary arrival-count distributions. Our approach differs in two important respects. First, we consider general substitutable choice models, reducing the computational component to the joint assortment and customization (JAC) problem. Second, and more importantly, our algorithm does not dynamically optimize assortments during the selling horizon. Instead, it selects the initial inventory and a fixed assortment for each customer type, subsequently offering all products from that assortment that remain in stock. Despite this restriction, we obtain approximation guarantees of $0.486\alpha_{\mathrm{JAC}}$ for deterministic or Poisson arrivals and $0.4328\alpha_{\mathrm{JAC}}$ for IFR arrivals, measured against the optimal fully adaptive DAP policy. For MNL choice, \citet{udwani2025submodular} provide a JAC approximation arbitrarily close to $1/2$, yielding guarantees approaching $0.243$ and $0.2164$, respectively. Although these constants do not improve the specialized MNL guarantees of \citet{sun2025unified}, our results show that simple type-specific show-all policies, combined with appropriately selected initial inventories, achieve provable performance guarantees even when compared with policies that continually adapt their assortments to the evolving inventory state.

\begin{table}[t]
\centering
\caption{Selected constant-factor guarantees for dynamic assortment personalization.}
\label{tab:dap_literature}
\begingroup
\small
\setlength{\tabcolsep}{4pt}
\renewcommand{\arraystretch}{1.15}
\begin{tabular}{@{}
>{\raggedright\arraybackslash}m{0.245\textwidth}
>{\raggedright\arraybackslash}m{0.275\textwidth}
>{\centering\arraybackslash}m{0.215\textwidth}
>{\centering\arraybackslash}m{0.17\textwidth}@{}}
\toprule
\textbf{Paper} & \textbf{Choice model} & \textbf{Deterministic} & \textbf{IFR} \\
\midrule
\citet{bai2025coordinated} & MNL & $\frac14(1-1/e)$ & --- \\
\addlinespace[3pt]
\citet{sun2025unified} & MNL & $\frac12(1-1/e)$ & $1/4^{\dagger}$ \\
\addlinespace[3pt]
\textbf{This paper} & Substitutable & $0.486\alpha_{\mathrm{JAC}}$ & $0.4328\alpha_{\mathrm{JAC}}$ \\
\addlinespace[3pt]
\textbf{This paper} & MNL; $\alpha_{\mathrm{JAC}}=1/2$ & $0.243$ & $0.2164$ \\
\bottomrule
\end{tabular}
\par\medskip
\begin{minipage}{\textwidth}
\footnotesize
\textit{Notes.} $\alpha_{\mathrm{JAC}}$ is the approximation factor
for joint assortment and customization, not cardinality-constrained
assortment optimization. The MNL instantiation uses the
$1/2-\varepsilon$ oracle of \citet{udwani2025submodular}; arbitrarily
small losses are suppressed. $\dagger$ The $1/4$ result holds for
arbitrary arrival-count distributions. Our Poisson and IFR results assume
i.i.d. customer types independent of $T$; deterministic arrivals permit
period-dependent type probabilities. A dash indicates that no
corresponding result is listed.
\end{minipage}
\endgroup
\end{table}

\subsection{Contributions}
\label{subsec:contributions}

\paragraph{A unified static-to-dynamic approximation framework.}
For both SA and DAP, we formulate a two-constraint fluid relaxation
that separates static product selection from stochastic inventory
implementation. Its columns are assortments for SA and customized
assortment configurations with a common catalog for DAP. An
approximate static oracle yields a solution supported on at most two
columns. We construct a feasible fractional inventory for each
supporting column and evaluate it using a separable lower bound on
expected truncated demand. Dependent rounding
\citep{gandhi2006dependent} preserves this bound in expectation,
because expected truncated demand is affine between adjacent integer
inventory levels. The analysis does not require the stochastic
revenue function to be concave or submodular.

\paragraph{Improved show-all guarantees and sharp fluid comparisons.}
For an $\alpha_{\mathrm{SA}}$-approximation to cardinality-constrained
assortment optimization, we obtain guarantees of
$0.486\alpha_{\mathrm{SA}}$ for deterministic/Poisson arrivals and
$0.4328\alpha_{\mathrm{SA}}$ for IFR arrivals. The deterministic/Poisson
analysis uses a Poisson truncated-demand surrogate and a sharper
conditional Cauchy--Schwarz bound. The IFR analysis replaces it with a
geometric surrogate and optimally balances the two fluid constraints.
With an exact oracle, the IFR factor can be made arbitrarily close to
$1-w\approx0.4328567$, where $we^w=1$; a matching instance makes this
the sharp uniform factor relative to the fluid LP. A two-product MNL
instance gives a corresponding upper limit of $1/2$ on the fluid
comparison for deterministic arrivals. These limits concern the
relaxation, not approximation hardness relative to the stochastic
optimum.

\paragraph{Personalization without dynamic assortment reoptimization.}
We extend our framework to DAP through a configuration-based fluid relaxation that jointly selects initial inventory and an assortment for each customer type. An $\alpha_{\mathrm{JAC}}$-approximation for JAC yields guarantees of $0.486\alpha_{\mathrm{JAC}}$ for deterministic or Poisson arrivals and $0.4328\alpha_{\mathrm{JAC}}$ for IFR arrivals. A notable feature of our algorithm is its simplicity: after the initial planning stage, it implements a type-specific show-all policy, offering every available product from a predetermined assortment for each arriving customer type. In particular, the algorithm never strategically removes an in-stock product or reoptimizes assortments in response to stockouts. Nevertheless, its performance guarantees hold relative to the optimal adaptive DAP policy, which may modify assortments based on customer types, remaining inventory, and the entire purchase history. Our results therefore demonstrate that joint optimization of initial inventory and type-specific assortments can recover a substantial guaranteed fraction of the benefits of fully dynamic assortment optimization, without requiring dynamic assortment decisions.

\section{Problem Formulation and Main Results}
\label{sec:problem_formulation}

We study the joint optimization of initial inventory and assortment
decisions under general customer choice models. The decision process
unfolds in two stages. First, the decision maker (DM) selects the
initial inventory levels of a collection of products, subject to a
capacity constraint. Subsequently, customers arrive sequentially
and make purchase decisions from the assortments offered to them.
We consider two operating settings: \emph{show-all} (SA) inventory
selection, in which every in-stock product must be offered to each
customer, and \emph{dynamic assortment personalization} (DAP), in
which the DM observes the arriving customer's type and may customize
the assortment offered from the available inventory. We first
introduce the notation and assumptions shared by both settings,
then formalize each problem and state our main approximation
results.

\subsection{Preliminaries}
\label{subsec:preliminaries}

\paragraph{Products and customer choice.}
Let $\mathcal{N}=\{1,\ldots,n\}$ denote the collection of products,
where each unit of product $i\in\mathcal{N}$ generates revenue
$r_i\geq0$ when purchased. The no-purchase option is indexed by
$0$, with $r_0=0$.

Customers make purchase decisions according to a general discrete
choice model. For any assortment $S\subseteq\mathcal{N}$, let
$\pi(i,S)$ denote the probability that a customer purchases
product $i$ when offered $S$. We set $\pi(i,S)=0$ for $i\notin S$
and define $\pi(0,S)=1-\sum_{i\in S}\pi(i,S)$ as the no-purchase
probability. In the personalized setting, customers belong to
types $k\in[K]$, each characterized by potentially different
choice probabilities $\pi^k(i,S)$. We use $\pi(i,S)$ when the
choice model is common to all customers, as in the SA setting,
and $\pi^k(i,S)$ when distinguishing customer types is necessary.
Choice probabilities are stationary within each type, and
customer purchase decisions are independent across arrivals,
conditional on the offered assortments and customer types.

We impose the following standard substitutability assumption,
which requires that removing products from an assortment cannot
decrease the purchase probability of a product that remains
available.

\begin{assumption}[Substitutability]
\label{ass:substitutability}
For any assortments $A\subseteq S\subseteq\mathcal{N}$ and product
$i\in A$, we have $\pi(i,A)\geq\pi(i,S)$. In the personalized
setting, the same condition holds for every customer type
$k\in[K]$, i.e., $\pi^k(i,A)\geq\pi^k(i,S)$.
\end{assumption}

This assumption is satisfied by a broad class of choice models,
including MNL, mixed-MNL, and ranking-based models. Importantly,
we do not impose any particular parametric structure on the
choice probabilities.

\paragraph{Inventory and customer arrivals.}
Before the selling horizon begins, the DM selects an initial
inventory vector $\boldsymbol{c}=(c_i)_{i\in\mathcal{N}}$, where
$c_i\in\mathbb{Z}_+$ denotes the number of units stocked for
product $i$. The total number of units stocked cannot exceed
an integer capacity $B\geq1$. Thus, the set of feasible initial
inventory vectors is
\[
\mathcal{C}_B
=
\left\{
\boldsymbol{c}\in\mathbb{Z}_+^n:
\sum_{i\in\mathcal{N}}c_i\leq B
\right\}.
\]
Inventory cannot be replenished during the selling horizon, and
each purchase permanently consumes one unit of the corresponding
product.

Let $T$ denote the total number of customer arrivals, with
$\tau=\mathbb{E}[T]$. We consider three arrival settings:
(i) $T$ is a known deterministic integer; (ii) $T$ is Poisson
distributed with known mean $\tau$; or (iii) $T$ follows a
nonnegative integer-valued increasing-failure-rate (IFR)
distribution with finite mean $\tau$. In the latter two settings,
the realization of $T$ is not known to the DM in advance.
The arrival count is independent of customer preferences.
Additional assumptions concerning the distribution of customer
types in DAP are introduced in Section~\ref{subsec:dap_setting}.

Customers are indexed by $t\in[T]$. Let
$\boldsymbol{c}_t$ denote the remaining inventory immediately
before customer $t$ arrives, with
$\boldsymbol{c}_1=\boldsymbol{c}$. In either operating setting,
the DM offers an assortment
$S_t\subseteq\{i\in\mathcal{N}:c_{i,t}>0\}$, and the customer
selects an option $j_t\in S_t\cup\{0\}$. Inventory then evolves
according to
$c_{i,t+1}=c_{i,t}-\mathbf{1}\{j_t=i\}$
for every $i\in\mathcal{N}$.

For any fixed initial inventory and admissible assortment policy,
expected revenue is obtained by taking the expectation of
$\sum_{t=1}^T r_{j_t}$ over customer arrivals, types, and purchase
decisions. We distinguish the two operating settings below.

\subsection{Show-All Inventory Selection}
\label{subsec:sa_setting}

\paragraph{Problem formulation.}
In the show-all (SA) setting, the DM offers every in-stock
product to each arriving customer. Thus, the assortment offered
in period $t$ is
$
S_t=\{i\in\mathcal{N}:c_{i,t}>0\}.
$
The DM makes no assortment decisions during the selling horizon;
the displayed assortment evolves endogenously as products stock
out. In particular, the offered assortments form a nested
sequence.

For a given initial inventory vector $\boldsymbol{c}$, let
$
\mathrm{Rev}(\boldsymbol{c})
=
\expar{\sum_{t=1}^T r_{j_t}}
$
denote the expected revenue generated under the show-all policy.
The dependence of $j_t$ on the initial inventory vector is
suppressed to simplify notation. The SA inventory selection
problem is
\begin{equation}
\label{eqn:SA}
\tag{OPT-SA}
\operatorname{OPT}_{\mathrm{SA}}
=
\max_{\boldsymbol{c}\in\mathcal{C}_B}
\mathrm{Rev}(\boldsymbol{c}).
\end{equation}

\paragraph{Computational assumption.}
Our approach relies on the cardinality-constrained assortment
optimization problem. For any reward vector
$\boldsymbol{\rho}\in\mathbb{Q}^n$, define
\begin{equation}
\label{eqn:cardinality_assortment}
\operatorname{AO}_B(\boldsymbol{\rho})
=
\max_{\substack{S\subseteq\mathcal{N}:\\ |S|\leq B}}
\sum_{i\in S}\rho_i\pi(i,S).
\end{equation}
We assume access to an algorithm that approximates this problem
within factor $\alpha_{\mathrm{SA}}\in(0,1]$ in polynomial time,
for any modified reward vector. The exact-oracle case corresponds
to $\alpha_{\mathrm{SA}}=1$.

The role of this oracle is to solve a fluid relaxation of
\eqref{eqn:SA} through dual separation. We denote the optimal
value of this relaxation, introduced in
Section~\ref{subsec:step_1}, by
$\operatorname{OPT}_{\mathrm{FL}}$. As established by
\citet{fu2026joint},
$\operatorname{OPT}_{\mathrm{FL}}\geq
\operatorname{OPT}_{\mathrm{SA}}$.

\paragraph{Deterministic and Poisson arrivals.}
Our first result substantially improves the approximation
guarantee obtainable from this relaxation when the number of
customers is deterministic or Poisson distributed.

\begin{theorem}[Deterministic and Poisson arrivals]
\label{thm:main_approximation}
Suppose Assumption~\ref{ass:substitutability} holds and $T$ is
either deterministic or Poisson distributed. If
\eqref{eqn:cardinality_assortment} can be solved optimally in
polynomial time, then there exists a randomized polynomial-time
algorithm that selects an integral inventory vector
$\widehat{\boldsymbol{c}}\in\mathcal{C}_B$ almost surely and
satisfies
\[
\mathbb{E}_{\omega}
\left[
\mathrm{Rev}(\widehat{\boldsymbol{c}}(\omega))
\right]
\geq
0.486\,\operatorname{OPT}_{\mathrm{FL}}
\geq
0.486\,\operatorname{OPT}_{\mathrm{SA}}.
\]
Here, $\omega$ denotes the initial-inventory rounding randomness.
\end{theorem}

When only an $\alpha_{\mathrm{SA}}$-approximation for
\eqref{eqn:cardinality_assortment} is available, the guarantee
becomes $0.486\alpha_{\mathrm{SA}}$, with an arbitrarily small
additional loss that can be absorbed into the stated numerical
constant.

The analysis also establishes that this guarantee is nearly
tight relative to the fluid relaxation. In particular, we
construct a sequence of MNL instances with deterministic arrivals
for which
$
\operatorname{OPT}_{\mathrm{SA}}/
\operatorname{OPT}_{\mathrm{FL}}
\rightarrow1/2.
$
Thus, no algorithm can guarantee a factor strictly larger than
$1/2$ relative to this fluid benchmark on all deterministic-arrival
instances. Our $0.486$ guarantee is therefore close to the
strongest possible factor obtainable through a direct comparison
with this relaxation.

\paragraph{IFR arrivals.}
We next consider the more general setting in which $T$ follows
an IFR distribution. The main difficulty is that hypothetical
product demand need not be dominated in convex order by a
Poisson random variable with the same mean. We overcome this
difficulty through a geometric-demand comparison and a modified
inventory allocation, obtaining the following result.

\begin{theorem}[IFR arrivals]
\label{thm:ifr_approximation}
Suppose Assumption~\ref{ass:substitutability} holds and $T$ is
a nonnegative integer-valued IFR random variable with finite
mean. If \eqref{eqn:cardinality_assortment} can be solved
optimally in polynomial time, then there exists a randomized
algorithm with expected polynomial running time that selects
an integral inventory vector
$\widehat{\boldsymbol{c}}\in\mathcal{C}_B$ almost surely and
satisfies
\[
\mathbb{E}_{\omega}
\left[
\mathrm{Rev}(\widehat{\boldsymbol{c}}(\omega))
\right]
\geq
0.4328\,\operatorname{OPT}_{\mathrm{FL}}
\geq
0.4328\,\operatorname{OPT}_{\mathrm{SA}}.
\]
Moreover, the factor can be made arbitrarily close to
$\gamma_{\mathrm{IFR}}=1-w\approx0.4328567$, where
$w\in(0,1)$ is the unique solution to $we^w=1$.
\end{theorem}

With an $\alpha_{\mathrm{SA}}$-approximation for
\eqref{eqn:cardinality_assortment}, the guarantee becomes
$0.4328\alpha_{\mathrm{SA}}$, again allowing an arbitrarily small
additional computational loss.

The IFR guarantee is essentially tight relative to the fluid
relaxation. We construct a sequence of instances with geometric
arrival counts and substitutable choice probabilities for which
\[
\frac{\operatorname{OPT}_{\mathrm{SA}}}
{\operatorname{OPT}_{\mathrm{FL}}}
\longrightarrow
\gamma_{\mathrm{IFR}}.
\]
Together with Theorem~\ref{thm:ifr_approximation}, this shows
that $\gamma_{\mathrm{IFR}}$ is the sharp uniform comparison factor
for the fluid relaxation under IFR arrivals.

These results apply to a broad class of substitutable choice
models. For example, cardinality-constrained assortment
optimization is polynomially solvable under MNL choice, giving
$\alpha_{\mathrm{SA}}=1$. Under general ranking-based choice with
uniform revenues, the greedy algorithm achieves
$\alpha_{\mathrm{SA}}=1-1/e$. Thus, the corresponding guarantees
are $0.486$ and $0.4328$ for MNL, and approximately $0.3072$
and $0.2736$ for uniform-revenue ranking-based choice, under
deterministic/Poisson and IFR arrivals, respectively.

\subsection{Dynamic Assortment Personalization}
\label{subsec:dap_setting}

\paragraph{Problem formulation.}
We next consider the dynamic assortment personalization (DAP)
problem. In this setting, customers belong to $K$ observable
types, indexed by $k\in[K]$, and a type-$k$ customer makes
purchase decisions according to $\pi^k(i,S)$.

When $T$ is deterministic, the customer arriving in period $t$
has type $k$ with probability $\lambda_{kt}$, where these
probabilities may vary across periods. When $T$ is Poisson or
IFR distributed, customer types are independent and identically
distributed with probabilities $\lambda_k$, independently of
$T$. Customer types and preference realizations are independent
across arrivals.

After observing the arriving customer's type, the DM may offer
any subset of the currently in-stock products. Moreover, the
assortment decision may depend on the entire preceding purchase
history and the remaining inventory vector. Thus, unlike SA,
the DM may strategically withhold products that remain available
and dynamically modify assortments as inventory is depleted.

Let $\Sigma_{\mathrm{DAP}}(\boldsymbol{c})$ denote the collection
of admissible nonanticipative personalized assortment policies
under initial inventory $\boldsymbol{c}$. For
$\sigma\in\Sigma_{\mathrm{DAP}}(\boldsymbol{c})$, let
$
\mathrm{Rev}^{\mathrm{DAP}}(\boldsymbol{c},\sigma)
=
\expar{\sum_{t=1}^T r_{j_t}}
$
denote the expected revenue under policy $\sigma$, where the
expectation includes customer arrivals, types, purchase decisions,
and any policy randomization. The DAP problem is
\begin{equation}
\label{eqn:DAP}
\tag{OPT-DAP}
\operatorname{OPT}_{\mathrm{DAP}}
=
\max_{\boldsymbol{c}\in\mathcal{C}_B}
\sup_{\sigma\in\Sigma_{\mathrm{DAP}}(\boldsymbol{c})}
\mathrm{Rev}^{\mathrm{DAP}}(\boldsymbol{c},\sigma).
\end{equation}

\paragraph{Joint assortment and customization.}
Our approach to DAP relies on a static optimization problem that
jointly selects a common product catalog and a customized
assortment for each customer type.

A \emph{configuration}
$\boldsymbol{A}=(A_1,\ldots,A_K)$ specifies an assortment $A_k$
for each type $k$. Its common product catalog is
$I_{\boldsymbol{A}}=\bigcup_{k=1}^K A_k$. Define the set of
feasible configurations by
\[
\mathfrak{A}_B
=
\left\{
\boldsymbol{A}=(A_1,\ldots,A_K):
A_k\subseteq\mathcal{N}\ \forall k,\quad
\left|\bigcup_{k=1}^K A_k\right|\leq B
\right\}.
\]
The common-catalog constraint is important: although different
customer types may be offered different assortments, they share
the same initial inventory, and a product may be purchased by
several types.

Let $\tau_k$ denote the expected number of type-$k$ arrivals.
For any modified reward vector
$\boldsymbol{\rho}\in\mathbb{Q}^n$, define the joint assortment
and customization (JAC) problem as
\begin{equation}
\label{eqn:JAC}
\operatorname{JAC}_B(\boldsymbol{\rho})
=
\max_{\boldsymbol{A}\in\mathfrak{A}_B}
\sum_{k=1}^K\tau_k
\sum_{i\in A_k}\rho_i\pi^k(i,A_k).
\end{equation}
We assume that this problem admits an
$\alpha_{\mathrm{JAC}}$-approximation in polynomial time for
every rational reward vector.

\paragraph{Main result.}
Our DAP results are obtained by extending the fluid relaxation
and inventory construction developed for SA to configurations
of personalized assortments. The resulting algorithms have a
particularly simple operating structure. They select both the
initial inventory and one assortment $A_k$ for each customer
type before the selling horizon begins. Whenever a type-$k$
customer arrives, the DM offers every product in $A_k$ that
remains in stock. Thus, the assortment offered to each type
changes only through stockouts, and the DM never strategically
removes an available product or reoptimizes assortments during
the selling horizon.

Despite this restriction, our performance guarantees are measured
against the optimal DAP policy, which may dynamically adapt
assortments to customer types, remaining inventory, and purchase
history.

\begin{theorem}[Approximation for DAP]
\label{thm:dap_approximation}
Suppose Assumption~\ref{ass:substitutability} holds for every
customer type and that \eqref{eqn:JAC} admits a polynomial-time
$\alpha_{\mathrm{JAC}}$-approximation for every rational reward
vector. Then there exists a randomized polynomial-time algorithm
that selects an initial inventory vector
$\widehat{\boldsymbol{c}}\in\mathcal{C}_B$ almost surely and a
fixed assortment $A_k$ for each customer type $k$, such that
offering $A_k$ intersected with the in-stock products to every
type-$k$ customer achieves the following guarantees:
\begin{enumerate}[label=(\roman*)]
    \item If $T$ is deterministic or Poisson distributed, the
    expected revenue is at least
    $0.486\alpha_{\mathrm{JAC}}\operatorname{OPT}_{\mathrm{DAP}}$.
    \item If $T$ is IFR distributed, the expected revenue is at
    least
    $0.4328\alpha_{\mathrm{JAC}}
    \operatorname{OPT}_{\mathrm{DAP}}$.
\end{enumerate}

\end{theorem}

Theorem~\ref{thm:dap_approximation} shows that a substantial
guaranteed fraction of the optimal adaptive DAP revenue can be
obtained through a type-specific show-all policy, provided that
the initial inventory and type-specific assortments are selected
jointly. In particular, no dynamic assortment reoptimization is
required after the selling horizon begins. This illustrates how
coordinating initial inventory and customer-specific assortment
decisions can reduce the need for more sophisticated dynamic
assortment control.

For MNL choice, a polynomial-time approximation algorithm for
JAC achieves a factor arbitrarily close to $1/2$. Consequently,
Theorem~\ref{thm:dap_approximation} yields DAP guarantees
approaching $0.243$ for deterministic or Poisson arrivals and
$0.2164$ for IFR arrivals. More generally, any improvement in
the approximation factor for JAC translates directly into an
improved DAP guarantee through our framework.

\section{Proof of Theorem~\ref{thm:main_approximation}} \label{sec:proof_main_thm}

The proof of Theorem~\ref{thm:main_approximation} proceeds in three main steps, which we summarize below. All proofs from this section can be found in Appendix~\ref{app:proofs}.

\paragraph{Step 1: A fluid upper bound (Section~\ref{subsec:step_1}).} 
We begin by introducing a fluid linear programming relaxation that allocates the expected customer arrivals across assortments, subject to constraints on total expected sales and the cardinality of each assortment. At this stage, we do not require an integral inventory decision; the relaxation accounts only for the expected consumption of inventory. We establish that every feasible inventory vector induces a feasible fluid solution with the same expected revenue, and hence the optimal fluid objective upper-bounds $\operatorname{OPT}_{\mathrm{SA}}$. Following~\cite{fu2026joint}, we solve this relaxation through its dual, whose constraints are indexed by assortments. Separating these constraints amounts to solving a cardinality-constrained assortment optimization problem with modified product revenues. Thus, our tractability assumption provides a separation oracle for generating violated dual constraints and solving the fluid LP in polynomial time. Moreover, since the fluid LP has only two resource constraints, we can select an optimal basic solution supported on at most two assortments.

\paragraph{Step 2: Constructing fractional inventory  (Section~\ref{subsec:step_2}).}
For each assortment $S$ in the support of the optimal fluid solution, we construct a fractional inventory vector that stocks only products in $S$. We continue to set aside integrality in this step, while requiring that the total inventory does not exceed $B$. The construction balances the inventory needed to stock a product against the inventory needed to serve its expected demand. To evaluate the resulting allocation, we introduce a separable expected-revenue lower bound based on the demand generated by offering $S$ throughout the horizon. Substitutability connects this auxiliary demand model to the show-all process, while bounds on expected truncated demand guide the choice of fractional inventory.

Let $D_S$ and $V_S$ denote, respectively, the expected total demand and revenue from offering $S$ throughout the horizon without inventory constraints, and write $d_S=D_S/B$. We show that the fractional inventory vector has an associated revenue lower bound of at least $L_S=V_S/(C+d_S)$, where $C \approx 1.057$. We then select the assortment with the largest value of $L_S$. The two constraints in the fluid LP imply that this certificate is at least a $1/(1+C) \approx 0.486$ fraction of the optimal fluid objective. Thus, this step identifies a capacity-feasible fractional inventory vector with the desired performance certificate, leaving only its conversion into an integral stocking decision.

\paragraph{Step 3: Randomized rounding  (Section~\ref{subsec:step_3}).}
Finally, we convert the selected fractional inventory vector into an integral one using the dependent-rounding procedure of~\cite{gandhi2006dependent}. Applied to the fractional parts of the inventory levels, with a dummy coordinate accounting for unused capacity, this procedure rounds each coordinate to its floor or ceiling, preserves its expectation, and satisfies the total inventory constraint in every realization. The key observation is that the expected truncated-demand function is affine between consecutive integer inventory levels, because customer demand is integer-valued. Consequently, the rounding preserves the expected value of the separable revenue lower bound from Step 2. Since this lower bound is dominated by the expected show-all revenue for every realized integral inventory vector, the rounded solution retains the certificate $L_S$ in expectation. Combining the three steps therefore yields a $1/(1+C)$-approximation without requiring any assortment decisions after the initial inventory is selected.

\subsection{Step 1: A fluid upper bound} \label{subsec:step_1}

We begin with the fluid relaxation of~\cite{fu2026joint}, which allocates
the expected customer arrivals across assortments while accounting for
the total inventory capacity in expectation. At this stage, we do not
select an inventory vector or impose integrality on expected sales.
Instead, the relaxation retains two implications of the inventory
constraint: the expected number of units sold cannot exceed $B$, and
every offered assortment contains at most $B$ products. We first
establish that the resulting linear program upper-bounds
$\operatorname{OPT}_{\mathrm{SA}}$ and then describe how to solve it
using a cardinality-constrained assortment optimization oracle.

\paragraph{The fluid relaxation.}
Let $\tau=\expar{T}$ denote the expected number of customer arrivals,
and define
$\mathcal{S}_B=\{S\subseteq\mathcal{N}:|S|\leq B\}$.
We assume $\tau>0$, since the problem is trivial otherwise.
For every assortment $S\in\mathcal{S}_B$, let
\[
\mu_i^S=\tau\pi(i,S),
\qquad
D_S=\sum_{i\in S}\mu_i^S,
\qquad
V_S=\sum_{i\in S}r_i\mu_i^S.
\]
Thus, $\mu_i^S$ is the expected demand for product $i$ if assortment
$S$ is offered to every arriving customer without inventory
restrictions, while $D_S$ and $V_S$ are the corresponding expected
total demand and revenue. For each $S\in\mathcal{S}_B$, introduce a nonnegative variable $f_S$,
representing the expected number of customers offered $S$, normalized
by $\tau$. The fluid relaxation is
\begin{equation}
\label{eqn:fluid_lp}
\tag{FLUID-LP}
\begin{aligned}
\operatorname{OPT}_{\mathrm{FL}}
=
\max_{\boldsymbol{f}\geq\boldsymbol{0}}\quad
&\sum_{S\in\mathcal{S}_B}f_SV_S\\
\text{s.t.}\quad
&\sum_{S\in\mathcal{S}_B}f_SD_S\leq B,\\
&\sum_{S\in\mathcal{S}_B}f_S\leq1.
\end{aligned}
\end{equation}
The first constraint limits total expected sales, while the second
limits the expected number of customers allocated to assortments.
Unlike the original problem,~\eqref{eqn:fluid_lp} does not require
the assortment allocations to arise from an integral starting inventory
vector or a stockout-driven sequence of offered assortments.
By Proposition~1 of~\cite{fu2026joint}, its optimal value satisfies
$\operatorname{OPT}_{\mathrm{FL}}\geq\operatorname{OPT}_{\mathrm{SA}}$.
This upper-bound property holds for arbitrary choice models and any
arrival count with finite mean that is independent of customer
preferences; substitutability and the distributional assumptions on
$T$ enter our subsequent inventory construction.

\paragraph{Solving the fluid relaxation.}
We also follow~\cite{fu2026joint} to solve~\eqref{eqn:fluid_lp}.
Although the primal has exponentially many assortment variables,
its dual has only two nonnegative variables, $\lambda$ and $\eta$,
associated with the expected-sales and customer-allocation constraints,
respectively. The dual constraints are
$\eta+\lambda D_S\geq V_S$ for every $S\in\mathcal{S}_B$.
Separating these constraints requires solving
\[
\max_{S\in\mathcal{S}_B}\{V_S-\lambda D_S\}
=
\tau
\max_{\substack{S\subseteq\mathcal{N}\\ |S|\leq B}}
\sum_{i\in S}(r_i-\lambda) \cdot \pi(i,S).
\]
Thus, a cardinality-constrained assortment optimization oracle with
modified rewards $\rho_i=r_i-\lambda$ identifies a violated dual
constraint whenever the displayed value exceeds $\eta$, and otherwise
certifies feasibility. This provides a separation oracle for dual
constraint generation. Under our computational assumptions,
Proposition~6 of~\cite{fu2026joint} establishes polynomial-time
solvability of the fluid LP through the ellipsoid method.

\paragraph{An optimal solution with small support.}
Since~\eqref{eqn:fluid_lp} has only two resource constraints, it
admits an optimal basic solution supported on at most two assortments,
as noted in Section~4.1 of~\cite{fu2026joint}. Fix such a solution
$\boldsymbol{f}^{\star}$, and let
$\mathcal{S}^{\star}
=\{S\in\mathcal{S}_B:f_S^{\star}>0\}$
denote its support. Then $|\mathcal{S}^{\star}|\leq2$ and
\[
\sum_{S\in\mathcal{S}^{\star}}f_S^{\star}V_S
=
\operatorname{OPT}_{\mathrm{FL}},
\qquad
\sum_{S\in\mathcal{S}^{\star}}f_S^{\star}D_S\leq B,
\qquad
\sum_{S\in\mathcal{S}^{\star}}f_S^{\star}\leq1.
\]
In the next step, we construct a capacity-feasible fractional inventory
vector for each $S\in\mathcal{S}^{\star}$ and use these inequalities
to compare the strongest resulting revenue certificate with
$\operatorname{OPT}_{\mathrm{FL}}$.

\subsection{Step 2: Constructing fractional inventory} \label{subsec:step_2}

Consider an arbitrary optimal fluid solution $\boldsymbol{f}^{\star}$ obtained in
Step 1, with support $\mathcal{S}^{\star}$. For each supporting
assortment $S$, we construct a capacity-feasible fractional inventory
vector that stocks only products in $S$. We then associate a revenue
certificate with each candidate and select the one with the largest
certificate. Integrality is not imposed in this step; the selected
fractional vector will be converted into an integral inventory
decision in Step 3.

\paragraph{The inventory construction.}
For $\beta\in(0,1)$, define
\begin{equation}
\label{eqn:inventory_slope}
b(\beta)
=
\begin{cases}
\beta,
&0<\beta\leq\frac12,\\[2mm]
\displaystyle
\frac{\beta}{2}+\frac{1}{8(1-\beta)},
&\frac12\leq\beta<1,
\end{cases}
\end{equation}
noting that the two expressions agree at $\beta=1/2$. Let $C>1$ be a fixed
rational constant satisfying
\begin{equation}
\label{eqn:admissible_constant}
\max_{\beta \in [0,1]}\frac{1}{\beta}
-
\frac{1-\beta}{b(\beta)}
\leq C.
\end{equation}
It is straightforward to establish that $C=1.0574$ satisfies
\eqref{eqn:admissible_constant}: the left-hand side equals one for
$\beta\leq1/2$ and attains its global maximum of approximately
$1.057309$ at $\beta\approx0.844446$. We therefore fix $C=1.0574$
throughout the analysis, which will ultimately yeild an approximation factor of
$1/(1+C)>0.486$. For every $S\in\mathcal{S}^{\star}$, let $d_S=D_S/B$ and define
$\beta_S=1/(C+d_S)$. Construct the fractional inventory vector
$\boldsymbol{z}^S$ according to
\begin{equation*}
\label{eqn:fractional_inventory}
z_i^S
=
\begin{cases}
\beta_S+b(\beta_S)\cdot \mu_i^S,
&i\in S,\\
0,
&i\notin S.
\end{cases}
\end{equation*}
This allocation combines a baseline quantity $\beta_S$ for each
product with a quantity proportional to its expected demand.
The parameter $\beta_S$ determines the fraction of expected demand
that our subsequent analysis guarantees. The vector $\boldsymbol{z}^S$ satisfies the total capacity constraint.
Indeed, applying~\eqref{eqn:admissible_constant} at $\beta_S$ gives
$
d_S
=
\frac{1}{\beta_S}-C
\leq
\frac{1-\beta_S}{b(\beta_S)}.
$
Consequently, $\beta_S+b(\beta_S)d_S\leq1$, and thus
\[
\sum_{i\in\mathcal{N}}z_i^S
=
\beta_S\cdot |S|+b(\beta_S)\cdot D_S\\
\leq
B\bigl(\beta_S+b(\beta_S)\cdot d_S\bigr)
\leq B,
\]
where we use $|S|\leq B$. Thus, each candidate is supported on its
corresponding assortment and is feasible apart from the integrality
requirement.

\paragraph{An expected-revenue lower bound.}
To evaluate the candidates, we first introduce a lower bound on the
expected revenue of an integral inventory vector supported on a
fixed assortment. For $\mu\geq0$, let $X_\mu$ be a Poisson random
variable with mean $\mu$, and define
$
g_\mu(q)
=
\expar{\min\{q,X_\mu\}}
$
 for any $q\geq0$. For an assortment $S$ and a nonnegative vector $\boldsymbol{z}$
supported on $S$, let
$
\mathcal{R}(\boldsymbol{z})
=
\sum_{i\in S}r_i \cdot  g_{\mu_i^S}(z_i).
$
For integral inventory, this expression lower-bounds expected
show-all revenue. For fractional inventory, it provides an
analytical extension of the same lower bound.

\begin{lemma}[Revenue lower bound]
\label{lem:revenue_lower_bound}
Suppose Assumption~\ref{ass:substitutability} holds and $T$ is
deterministic or Poisson distributed. For every assortment $S$ and
integral inventory vector $\boldsymbol{c}$ supported on $S$,
\[
\mathrm{Rev}(\boldsymbol{c})
\geq
\mathcal{R}(\boldsymbol{c}).
\]
\end{lemma}
The function $\mathcal{R}(\cdot)$ allows us to evaluate fractional
inventory without defining a stochastic purchase process that
consumes fractional units. In the remainder of this step, we
establish a lower bound on
$\mathcal{R}(\boldsymbol{z}^S)$. Step 3 will show that
randomized rounding preserves this surrogate value in expectation,
thereby transferring the bound to a feasible integral inventory
decision.

\paragraph{A scalar truncated-demand bound.}
The choice of $b(\beta)$ in~\eqref{eqn:inventory_slope} is motivated
by the following inequality

\begin{lemma}
\label{lem:poisson_inventory_bound}
For every $\mu\geq0$ and $\beta\in(0,1)$,
\[
g_\mu\bigl(\beta+b(\beta)\cdot \mu\bigr)
\geq
\beta\mu.
\]
\end{lemma}
Applying Lemma~\ref{lem:poisson_inventory_bound} to each product in
$S$ gives
\begin{equation}
\label{eqn:assortment_revenue_certificate}
\begin{aligned}
\mathcal{R}(\boldsymbol{z}^S)
&=
\sum_{i\in S}
r_i\cdot  g_{\mu_i^S}
\bigl(\beta_S+b(\beta_S)\cdot \mu_i^S\bigr)\\
&\geq
\beta_S\sum_{i\in S}r_i\mu_i^S\\
&=
\beta_SV_S
=
\frac{V_S}{C+d_S}.
\end{aligned}
\end{equation}
 Thus, every supporting
assortment produces a capacity-feasible fractional inventory vector
with a corresponding certificate $L_S$.

\paragraph{Selecting the best candidate.} Let $L_S=\frac{V_S}{C+d_S}$ and choose
$
\widehat S
\in
\arg\max_{S\in\mathcal{S}^{\star}}L_S,
$
and retain the fractional inventory vector
$\boldsymbol{z}^{\widehat S}$.
The two resource constraints in the fluid LP imply
\begin{align}
\operatorname{OPT}_{\mathrm{FL}}
&=
\sum_{S\in\mathcal{S}^{\star}}f_S^\star V_S
\nonumber\\
&=
\sum_{S\in\mathcal{S}^{\star}}
f_S^\star \cdot 
\left(C+\frac{D_S}{B}\right) \cdot L_S
\nonumber\\
&\leq
L_{\widehat S} \cdot 
\left(
C\sum_{S\in\mathcal{S}^{\star}}f_S^\star
+
\frac{1}{B}
\sum_{S\in\mathcal{S}^{\star}}f_S^\star D_S
\right)
\nonumber\\
&\leq
(C+1)\cdot L_{\widehat S}.
\label{eqn:combine_assortment_certificates}
\end{align}
Combining~\eqref{eqn:assortment_revenue_certificate}
and~\eqref{eqn:combine_assortment_certificates}, we obtain
$
\mathcal{R}
(\boldsymbol{z}^{\widehat S})
\geq
L_{\widehat S}
\geq
\frac{\operatorname{OPT}_{\mathrm{FL}}}{1+C}
\geq
\frac{\operatorname{OPT}_{\mathrm{SA}}}{1+C}.
$
Thus, Step 2 identifies a capacity-feasible fractional inventory
vector whose revenue certificate is at least a $0.486$ fraction
of the optimal expected revenue.  In Step 3, we convert
$\boldsymbol{z}^{\widehat S}$ into an integral inventory vector
while preserving the surrogate in expectation, completing the
approximation guarantee.

\subsection{Step 3: Randomized rounding } \label{subsec:step_3}

The final step converts the fractional inventory vector selected in
Step 2 into a feasible integral inventory decision. Fix the selected
assortment $\widehat S$, and write
$\boldsymbol{z}=\boldsymbol{z}^{\widehat S}$.
Recall that $\boldsymbol{z}$ is supported on $\widehat S$ and
satisfies $\sum_i z_i\leq B$. We apply dependent rounding to obtain
an integral inventory vector that respects capacity in every
realization and preserves each coordinate's expectation. We then show that the
expected truncated-demand function is affine between consecutive
integer inventory levels, so this rounding also preserves the
surrogate value established in Step 2.

\paragraph{Rounding the fractional inventory.}
Fix the selected assortment $\widehat S$ and fractional inventory
vector $\boldsymbol{z}=\boldsymbol{z}^{\widehat S}$.
Let $\omega$ denote the randomness used by the dependent-rounding
procedure of~\cite{gandhi2006dependent}. The procedure produces
a random integral inventory vector
$\widehat{\boldsymbol{c}}=\widehat{\boldsymbol{c}}(\omega)$
satisfying
\begin{enumerate}[label=(\roman*)]
    \item $\widehat c_i\in
    \{\lfloor z_i\rfloor,\lceil z_i\rceil\}$ for every
    $i\in\mathcal{N}$, almost surely;
    \item $\Pr_{\omega}
    \{\widehat c_i=\lceil z_i\rceil\}
    =z_i-\lfloor z_i\rfloor$ for every $i\in\mathcal{N}$
    with $z_i\notin\mathbb{Z}$;
    \item $\sum_{i\in\mathcal{N}}\widehat c_i\leq B$
    almost surely.
\end{enumerate}
If $z_i$ is integral, then $\widehat c_i=z_i$ almost surely.
All probabilities and almost-sure statements above concern only
the rounding randomness.The next lemma shows that the rounding introduces no loss in the expected value of our separable revenue lower bound.
\begin{lemma}
\label{lem:rounding_preserves_surrogate}
The inventory vector $\widehat{\boldsymbol{c}}$ produced by the
dependent-rounding procedure satisfies
\[
\mathbb{E}_{\omega}
\left[
\mathcal{R}(\widehat{\boldsymbol{c}})
\right]
=
\mathcal{R}(\boldsymbol{z}),
\]
where the expectation is taken only over the rounding randomness.
\end{lemma}
Every realization of $\widehat{\boldsymbol{c}}$ is integral
and supported on $\widehat S$. Lemma~\ref{lem:revenue_lower_bound}
therefore gives
$\mathrm{Rev}(\widehat{\boldsymbol{c}}(\omega))
\geq
\mathcal{R}(\widehat{\boldsymbol{c}}(\omega))$
for every inventory vector produced by the rounding.
Here, the left-hand side already averages over customer arrivals
and choices; the inequality is not a pathwise comparison of
realized customer revenues.

\paragraph{Completing the approximation guarantee.}
Let $\xi$ denote the customer-arrival and preference randomness,
generated independently of the rounding randomness $\omega$.
For a fixed inventory vector $\boldsymbol{c}$, let
$\mathrm{Rev}(\boldsymbol{c};\xi)$ denote the realized show-all revenue,
so that
$\mathrm{Rev}(\boldsymbol{c})
=\mathbb{E}_{\xi}[\mathrm{Rev}(\boldsymbol{c};\xi)]$.
Our algorithm stocks $\widehat{\boldsymbol{c}}(\omega)$ and
subsequently offers every in-stock product. Its expected revenue
is therefore the leftmost quantity in the following display,
which averages over both the initial inventory randomization and
the subsequent customer arrivals and choices. Combining
Lemma~\ref{lem:rounding_preserves_surrogate} with the surrogate
bound from Step 2 and the certificate comparison
in~\eqref{eqn:combine_assortment_certificates}, we obtain
\[
\begin{aligned}
\mathbb{E}_{\omega,\xi}
\left[
\mathrm{Rev}(\widehat{\boldsymbol{c}}(\omega);\xi)
\right]
&=
\mathbb{E}_{\omega}
\left[
\rev(\widehat{\boldsymbol{c}}(\omega))
\right]\\
&\geq
\mathbb{E}_{\omega}
\left[
\mathcal{R}(\widehat{\boldsymbol{c}}(\omega))
\right]\\
&=
\mathcal{R}(\boldsymbol{z}^{\widehat S})\\
&\geq
L_{\widehat S}\\
&\geq
\frac{\operatorname{OPT}_{\mathrm{FL}}}{1+C}\\
&\geq
\frac{\operatorname{OPT}_{\mathrm{SA}}}{1+C}\\
&\geq
0.486\,\operatorname{OPT}_{\mathrm{SA}}.
\end{aligned}
\]
The first equality follows by conditioning on the rounding
randomness and then averaging over customer arrivals and choices.
The final inequality uses $C=1.0574$ and
$1/(1+C)=5000/10287>0.486$.
Thus, the algorithm's expected revenue is at least a $0.486$
fraction of the optimal expected revenue, while its inventory
capacity constraint is satisfied for every realization of the
rounding.

\paragraph{Computational complexity.}
Step 1 obtains an optimal fluid solution with at most two supporting
assortments in polynomial time under our static assortment
optimization assumption. Step 2 evaluates the choice probabilities
for these assortments and constructs their fractional inventory
vectors using rational arithmetic.  Finally, the rounding operates only on the fractional parts of the inventory coordinates and requires $O(n)$ pairwise updates; it does not enumerate the individual inventory units. Exact sampling of the
rational update probabilities can be implemented in expected
polynomial time. Thus, the complete algorithm has expected running
time polynomial in the input encoding.

\subsection{A Factor-Two Gap for the Fluid Relaxation}
\label{subsec:fluid_gap}

Before proceeding, we show that the fluid relaxation~\eqref{eqn:fluid_lp} can overestimate the optimal expected revenue by a factor approaching two. Thus, no algorithm whose performance is measured directly relative to $\operatorname{OPT}_{\mathrm{FL}}$ can obtain a uniform approximation guarantee strictly larger than $1/2$. The example holds even under the standard MNL choice model with only two products and a deterministic number of customer arrivals.

\begin{proposition}
\label{prop:fluid_gap}
There exists a sequence of MNL instances for which
\[
\frac{\operatorname{OPT}_{\mathrm{SA}}}
{\operatorname{OPT}_{\mathrm{FL}}}
\longrightarrow
\frac12.
\]
\end{proposition}
Proposition~\ref{prop:fluid_gap} implies that the factor $1/2$ is a fundamental upper bound on any uniform performance guarantee measured directly against $\operatorname{OPT}_{\mathrm{FL}}$. In particular, improving our $0.486$ guarantee all the way to $1/2$ would be essentially tight relative to this relaxation.

\section{Proof of Theorem~\ref{thm:ifr_approximation}}
\label{sec:proof_ifr}

We now establish the approximation guarantee for IFR-distributed
customer arrivals. Our approach follows the same three steps as
the proof of Theorem~\ref{thm:main_approximation}, with two
modifications to the fractional inventory construction. First,
we replace the Poisson truncated-demand lower bound with one
based on a geometric distribution. Second, we introduce separate
coefficients for the two constraints of the fluid LP, allowing
us to obtain a stronger approximation guarantee.

\paragraph{Step 1: A fluid upper bound
(Section~\ref{subsec:ifr_step_1}).}
We use the same fluid relaxation~\eqref{eqn:fluid_lp} introduced
in Section~\ref{subsec:step_1}. Its upper-bound property holds
for any distribution of $T$ with finite mean, including IFR
distributions, and its solution requires only the cardinality-
constrained assortment optimization oracle. As before, we fix
an optimal basic solution supported on at most two assortments.

\paragraph{Step 2: Constructing fractional inventory
(Section~\ref{subsec:ifr_step_2}).}
For each supporting assortment $S$, we construct a fractional
inventory vector supported on $S$ and associate it with a
separable revenue lower bound. The main difference is that
hypothetical product demand under IFR arrivals need not be
dominated in convex order by a Poisson random variable with
the same mean. Instead, we show that it is dominated by a
geometric random variable, which provides a lower bound on
expected truncated sales. Using this comparison, we develop
a new inventory allocation and obtain a certificate of the form
\[
L_S^{\mathrm{IFR}}
=
\frac{V_S}{C_0+C_1D_S/B},
\]
where $C_0=1.2148$ and $C_1=1.0955$.
We select the supporting assortment with the largest certificate
and show that its value is at least
$1/(C_0+C_1)>0.4328$ times the optimal fluid objective.

\paragraph{Step 3: Randomized rounding
(Section~\ref{subsec:ifr_step_3}).}
Finally, we apply the same dependent-rounding procedure as
in Section~\ref{subsec:step_3}. Since geometric demand is
integer-valued, its expected truncated-demand function is
affine between consecutive integer inventory levels. Thus,
the rounding preserves our new revenue surrogate in expectation,
while satisfying the capacity constraint almost surely.
Combining the three steps establishes
Theorem~\ref{thm:ifr_approximation}.

\subsection{Step 1: A fluid upper bound}
\label{subsec:ifr_step_1}

We continue to use the fluid relaxation~\eqref{eqn:fluid_lp},
which depends on the arrival distribution only through its
mean $\tau=\expar{T}$. Recall that, for each
$S\in\mathcal{S}_B$, we define
\[
\mu_i^S=\tau\pi(i,S),
\qquad
D_S=\sum_{i\in S}\mu_i^S,
\qquad
V_S=\sum_{i\in S}r_i\mu_i^S.
\]
As discussed in Section~\ref{subsec:step_1}, the upper-bound
property
$\operatorname{OPT}_{\mathrm{FL}}\geq
\operatorname{OPT}_{\mathrm{SA}}$
holds for arbitrary distributions of $T$ with finite mean.

Moreover, the same assortment optimization oracle allows us to
solve the fluid LP in polynomial time. Fix an optimal basic
solution $\boldsymbol{f}^{\star}$ with support
$\mathcal{S}^{\star}$, where
$|\mathcal{S}^{\star}|\leq2$ and
\[
\sum_{S\in\mathcal{S}^{\star}}f_S^\star V_S
=
\operatorname{OPT}_{\mathrm{FL}},
\qquad
\sum_{S\in\mathcal{S}^{\star}}f_S^\star D_S\leq B,
\qquad
\sum_{S\in\mathcal{S}^{\star}}f_S^\star\leq1.
\]
The remainder of the proof is devoted to constructing an
inventory vector whose expected revenue captures a constant
fraction of this fluid objective.

\subsection{Step 2: Constructing fractional inventory}
\label{subsec:ifr_step_2}

As in Section~\ref{subsec:step_2}, we construct a fractional
inventory vector for each assortment in $\mathcal{S}^{\star}$.
We first establish the geometric-demand comparison that replaces
the Poisson lower bound used previously.

\paragraph{A geometric revenue surrogate.}
For $\mu>0$, let $G_\mu$ denote a geometric random variable
on $\mathbb{Z}_+$ with mean $\mu$, so that
\[
\Pr(G_\mu=k)
=
\frac{1}{1+\mu}
\left(\frac{\mu}{1+\mu}\right)^k,
\qquad k\in\mathbb{Z}_+.
\]
For $\mu=0$, set $G_\mu=0$ almost surely. Define
$
g_\mu^{\mathrm{G}}(q)
=
\expar{\min\{q,G_\mu\}}
$
for $q\geq0$. For any assortment $S$ and nonnegative inventory
vector $\boldsymbol{z}$ supported on $S$, define
\[
\mathcal{R}^{\mathrm{G}}_S(\boldsymbol{z})
=
\sum_{i\in S}
r_i g_{\mu_i^S}^{\mathrm{G}}(z_i).
\]
As in the deterministic and Poisson settings, this surrogate
is defined for fractional inventory only as an analytical
extension of the expected truncated-demand lower bound.

\begin{lemma}[Revenue lower bound under IFR arrivals]
\label{lem:ifr_revenue_lower_bound}
Suppose Assumption~\ref{ass:substitutability} holds and $T$
is IFR distributed. For every assortment $S$ and integral
inventory vector $\boldsymbol{c}$ supported on $S$,
\[
\mathrm{Rev}(\boldsymbol{c})
\geq
\mathcal{R}^{\mathrm{G}}_S(\boldsymbol{c}).
\]
\end{lemma}

The proof, given in Appendix~\ref{app:proofs}, follows the
same fixed-assortment coupling as
Lemma~\ref{lem:revenue_lower_bound}. The new ingredient is
that a binomial random variable with an IFR-distributed
number of trials is dominated in convex order by a geometric
random variable with the same mean. Since truncation is
concave, this gives the desired lower bound.

\paragraph{A scalar truncated-demand bound.}
For $\beta\in(0,1)$, define
\begin{equation}
\label{eqn:ifr_inventory_slope}
b_{\mathrm{G}}(\beta)
=
\begin{cases}
-\log(1-\beta),
&0<\beta\leq\frac{7}{10},\\[2mm]
\displaystyle
\beta+\frac{1}{4(1-\beta)},
&\frac{7}{10}<\beta<1.
\end{cases}
\end{equation}
The two branches arise from different bounds on geometric
truncated demand. The first exploits the exact geometric
tail probabilities, while the second uses Cauchy--Schwarz
conditional on positive demand.

\begin{lemma}[Geometric truncated-demand bound]
\label{lem:ifr_inventory_bound}
For every $\mu\geq0$ and $\beta\in(0,1)$,
\[
g_\mu^{\mathrm{G}}
\bigl(\beta+b_{\mathrm{G}}(\beta)\cdot\mu\bigr)
\geq
\beta\mu.
\]
\end{lemma}

Unlike the Poisson case, we allow different coefficients for
the two constraints in the fluid relaxation. In particular,
we select constants $C_0,C_1>0$ satisfying
\begin{equation}
\label{eqn:ifr_admissible_constants}
\frac{1}{\beta}
-
C_1\frac{1-\beta}{b_{\mathrm{G}}(\beta)}
\leq C_0
\qquad\forall\beta\in(0,1).
\end{equation}

\begin{lemma}[Admissible coefficients]
\label{lem:ifr_admissible_constants}
The rational constants
$C_0=1.2148$ and $C_1=1.0955$
satisfy~\eqref{eqn:ifr_admissible_constants}.
Moreover, letting $w\in(0,1)$ be the unique solution to
$we^w=1$, the condition also holds for
\[
C_1^\star
=
\frac{w^2}{(1-w)^2(1+w)},
\qquad
C_0^\star
=
\frac{1}{1-w}-C_1^\star.
\]
These constants satisfy
$C_0^\star+C_1^\star=1/(1-w)$.
\end{lemma}

The proofs of Lemmas~\ref{lem:ifr_inventory_bound}
and~\ref{lem:ifr_admissible_constants} are given in
Appendix~\ref{app:proofs}. We use the rational constants
$C_0=1.2148$ and $C_1=1.0955$ in the algorithm, while the
limiting constants will be useful for establishing tightness
relative to the fluid relaxation.

\paragraph{The inventory construction.}
Consider any $S\in\mathcal{S}^{\star}$ with $D_S>0$, and let
$d_S=D_S/B$. Define
\[
\beta_S
=
\frac{1}{C_0+C_1d_S},
\qquad
\eta_S
=
\frac{1-\beta_S}{d_S}.
\]
Construct the fractional inventory vector
$\boldsymbol{z}^S$ according to
\begin{equation}
\label{eqn:ifr_fractional_inventory}
z_i^S
=
\begin{cases}
\beta_S+\eta_S\cdot\mu_i^S,
&i\in S,\\
0,
&i\notin S.
\end{cases}
\end{equation}
As before, the allocation consists of a baseline inventory
for every product, together with an amount proportional to
its expected demand. Assortments with $D_S=0$ also have
$V_S=0$ and can be discarded.

The vector $\boldsymbol{z}^S$ satisfies the total capacity
constraint. Indeed, since
$\eta_Sd_S=1-\beta_S$ and $|S|\leq B$,
\[
\begin{aligned}
\sum_{i\in\mathcal{N}}z_i^S
&=
\beta_S|S|+\eta_SD_S\\
&\leq
B(\beta_S+\eta_Sd_S)
=
B.
\end{aligned}
\]
Thus, each candidate is feasible apart from integrality.

\paragraph{An expected-revenue certificate.}
Applying~\eqref{eqn:ifr_admissible_constants} at $\beta_S$
and using
$C_1d_S=1/\beta_S-C_0$, we obtain
\[
d_S
\leq
\frac{1-\beta_S}{b_{\mathrm{G}}(\beta_S)},
\]
which implies
$\eta_S\geq b_{\mathrm{G}}(\beta_S)$.
By the monotonicity of $g_\mu^{\mathrm{G}}$ and
Lemma~\ref{lem:ifr_inventory_bound},
\[
\begin{aligned}
g_{\mu_i^S}^{\mathrm{G}}(z_i^S)
&\geq
g_{\mu_i^S}^{\mathrm{G}}
\bigl(\beta_S+b_{\mathrm{G}}(\beta_S)\mu_i^S\bigr)\\
&\geq
\beta_S\mu_i^S.
\end{aligned}
\]
Multiplying by $r_i$ and summing over products gives
\begin{equation}
\label{eqn:ifr_assortment_certificate}
\begin{aligned}
\mathcal{R}^{\mathrm{G}}_S(\boldsymbol{z}^S)
&=
\sum_{i\in S}
r_i g_{\mu_i^S}^{\mathrm{G}}(z_i^S)\\
&\geq
\beta_S\sum_{i\in S}r_i\mu_i^S\\
&=
\beta_SV_S
=
\frac{V_S}{C_0+C_1d_S}.
\end{aligned}
\end{equation}
Thus, every supporting assortment produces a capacity-feasible
fractional inventory vector with a corresponding revenue
certificate.

\paragraph{Selecting the best candidate.}
Let
$
L_S^{\mathrm{IFR}}
=
\frac{V_S}{C_0+C_1d_S}
$
and choose
$
\widehat S
\in
\arg\max_{S\in\mathcal{S}^{\star}}
L_S^{\mathrm{IFR}}.
$
Retain the fractional inventory vector
$\boldsymbol{z}^{\widehat S}$.

Using the two constraints of the fluid LP, we obtain
\begin{align}
\operatorname{OPT}_{\mathrm{FL}}
&=
\sum_{S\in\mathcal{S}^{\star}}f_S^\star V_S
\nonumber\\
&=
\sum_{S\in\mathcal{S}^{\star}}
f_S^\star
\left(C_0+C_1\frac{D_S}{B}\right)
L_S^{\mathrm{IFR}}
\nonumber\\
&\leq
L_{\widehat S}^{\mathrm{IFR}}
\left(
C_0\sum_{S\in\mathcal{S}^{\star}}f_S^\star
+
\frac{C_1}{B}
\sum_{S\in\mathcal{S}^{\star}}f_S^\star D_S
\right)
\nonumber\\
&\leq
(C_0+C_1)L_{\widehat S}^{\mathrm{IFR}}.
\label{eqn:ifr_combine_certificates}
\end{align}
Combining~\eqref{eqn:ifr_assortment_certificate}
and~\eqref{eqn:ifr_combine_certificates}, we conclude that
\[
\mathcal{R}^{\mathrm{G}}_{\widehat S}
(\boldsymbol{z}^{\widehat S})
\geq
L_{\widehat S}^{\mathrm{IFR}}
\geq
\frac{\operatorname{OPT}_{\mathrm{FL}}}{C_0+C_1}.
\]
Since
$C_0+C_1=2.3103$, we have
\[
\frac{1}{C_0+C_1}
=
\frac{10000}{23103}
>
0.4328.
\]
Thus, Step 2 identifies a capacity-feasible fractional inventory
vector whose geometric-demand revenue surrogate captures more
than $0.4328$ of the optimal fluid objective.

\subsection{Step 3: Randomized rounding}
\label{subsec:ifr_step_3}

The rounding procedure is identical to that in
Section~\ref{subsec:step_3}. Fix the selected assortment
$\widehat S$ and fractional inventory vector
$\boldsymbol{z}=\boldsymbol{z}^{\widehat S}$.
Apply the dependent-rounding procedure of
\cite{gandhi2006dependent} to obtain an integral inventory
vector $\widehat{\boldsymbol{c}}(\omega)$ satisfying the
three properties stated in Section~\ref{subsec:step_3}.
In particular,
$\widehat{\boldsymbol{c}}\in\mathcal{C}_B$ almost surely.

The key observation is that, since $G_\mu$ is integer-valued,
the function $g_\mu^{\mathrm{G}}(q)$ is affine between consecutive
integers. Therefore, the same argument as in the proof of
Lemma~\ref{lem:rounding_preserves_surrogate} gives
\[
\mathbb{E}_{\omega}
\left[
\mathcal{R}^{\mathrm{G}}_{\widehat S}
(\widehat{\boldsymbol{c}}(\omega))
\right]
=
\mathcal{R}^{\mathrm{G}}_{\widehat S}
(\boldsymbol{z}^{\widehat S}).
\]
Every realization of $\widehat{\boldsymbol{c}}$ is integral
and supported on $\widehat S$, so
Lemma~\ref{lem:ifr_revenue_lower_bound} also gives
\[
\mathrm{Rev}(\widehat{\boldsymbol{c}}(\omega))
\geq
\mathcal{R}^{\mathrm{G}}_{\widehat S}
(\widehat{\boldsymbol{c}}(\omega)).
\]

\paragraph{Completing the approximation guarantee.}
Combining these observations with
\eqref{eqn:ifr_assortment_certificate}
and~\eqref{eqn:ifr_combine_certificates}, we obtain
\[
\begin{aligned}
\mathbb{E}_{\omega}
\left[
\mathrm{Rev}(\widehat{\boldsymbol{c}}(\omega))
\right]
&\geq
\mathbb{E}_{\omega}
\left[
\mathcal{R}^{\mathrm{G}}_{\widehat S}
(\widehat{\boldsymbol{c}}(\omega))
\right]\\
&=
\mathcal{R}^{\mathrm{G}}_{\widehat S}
(\boldsymbol{z}^{\widehat S})\\
&\geq
L_{\widehat S}^{\mathrm{IFR}}\\
&\geq
\frac{\operatorname{OPT}_{\mathrm{FL}}}{C_0+C_1}\\
&\geq
\frac{\operatorname{OPT}_{\mathrm{SA}}}{C_0+C_1}\\
&\geq
0.4328\,\operatorname{OPT}_{\mathrm{SA}}.
\end{aligned}
\]
As before, $\mathrm{Rev}(\boldsymbol{c})$ already averages
over customer arrivals and choices, while the outer expectation
is taken only over the initial-inventory rounding randomness.

\paragraph{Computational complexity.}
Step 1 uses the same polynomial-time fluid optimization procedure
as Section~\ref{subsec:step_1}. Step 2 evaluates the demand
parameters and constructs fractional inventory using rational
arithmetic, since $C_0$ and $C_1$ are fixed rational constants.
Step 3 uses the same dependent-rounding procedure as before.
Thus, the overall algorithm has expected polynomial running time.

Finally, Lemma~\ref{lem:ifr_admissible_constants} establishes
that the admissibility condition also holds for
$C_0^\star,C_1^\star$, whose sum equals $1/(1-w)$.
Choosing rational coefficients arbitrarily close to these values
from above yields any fixed approximation factor strictly below
$1-w\approx0.4328567$, completing the proof of
Theorem~\ref{thm:ifr_approximation}.

\subsection{Tightness Relative to the Fluid Relaxation}
\label{subsec:ifr_fluid_gap}

We conclude by showing that the IFR approximation factor is
essentially tight relative to the fluid relaxation. Specifically,
we construct a sequence of instances for which the optimal
expected show-all revenue approaches only a
$(1-w)$-fraction of $\operatorname{OPT}_{\mathrm{FL}}$.
The example uses geometric arrivals and independent product
requests without substitution, demonstrating that this gap
does not require complicated choice behavior.

\begin{proposition}
\label{prop:ifr_fluid_gap}
There exists a sequence of instances with IFR-distributed
arrival counts and substitutable choice probabilities for which
\[
\frac{\operatorname{OPT}_{\mathrm{SA}}}
{\operatorname{OPT}_{\mathrm{FL}}}
\longrightarrow
1-w,
\]
where $w\in(0,1)$ is the unique solution to $we^w=1$.
\end{proposition}

The proof is provided in Appendix~\ref{app:proofs}. In the
constructed instances, the fluid relaxation obtains substantial
revenue from both a high-demand product and many products with
vanishingly small expected demand. An integral inventory decision,
however, must reserve at least one unit for each stocked
low-demand product, thereby reducing the capacity available to
the high-demand product. The resulting gap is amplified by the
variability of geometric demand.

Proposition~\ref{prop:ifr_fluid_gap}, together with
Theorem~\ref{thm:ifr_approximation}, establishes that
$1-w\approx0.4328567$ is the sharp uniform comparison factor
for this fluid relaxation under IFR arrivals. Consequently,
our $0.4328$ approximation guarantee is essentially tight
relative to the fluid benchmark.

\section{Proof of Theorem~\ref{thm:dap_approximation}}
\label{sec:proof_dap}

We now extend the preceding analysis to dynamic assortment
personalization. The proof follows the same three steps as the
SA analysis, replacing individual assortments with configurations
of type-specific assortments that share a common product catalog.
The main additional ingredients are a configuration-based fluid
upper bound and a revenue comparison for personalized assortment
plans. Once these are established, the fractional inventory
constructions and rounding arguments developed for SA apply
without modification.

\paragraph{Step 1: A configuration-based fluid upper bound
(Section~\ref{subsec:dap_step_1}).}
We formulate a fluid relaxation whose columns are configurations
$\boldsymbol{A}=(A_1,\ldots,A_K)$ satisfying
$|\bigcup_k A_k|\leq B$. We show that every feasible initial
inventory and adaptive personalized policy induces a feasible
fluid solution with the same expected revenue. The common-catalog
constraint is essential: different types may receive different
assortments, but these assortments must draw from products that
can be stocked together. An
$\alpha_{\mathrm{JAC}}$-approximation for the joint assortment
and customization problem yields a feasible fluid solution
with a corresponding approximation guarantee, supported on
at most two configurations.

\paragraph{Step 2: Constructing fractional inventory
(Section~\ref{subsec:dap_step_2}).}
For each supporting configuration, we construct an inventory
vector supported on its common catalog. Hypothetical demand
aggregates purchases across all customer types, allowing the
same inventory to serve multiple types. Under deterministic
or Poisson arrivals, we apply the Poisson inventory bound
from Section~\ref{subsec:step_2}; under IFR arrivals, we use
the geometric inventory bound from
Section~\ref{subsec:ifr_step_2}. These constructions give
configuration-level revenue certificates of the same form
as in the SA analysis. We select the configuration with the
largest certificate.

\paragraph{Step 3: Randomized rounding and implementation
(Section~\ref{subsec:dap_step_3}).}
We apply the dependent-rounding procedure from
Section~\ref{subsec:step_3} to the selected fractional inventory.
The implemented policy fixes one planned assortment for each
customer type before the selling horizon begins and subsequently
offers every in-stock product from that assortment. The same
interpolation argument preserves the revenue certificate in
expectation. Thus, the policy requires no dynamic assortment
reoptimization, although its guarantee is measured against
the optimal fully adaptive DAP policy.

All new lemma proofs are provided in
Appendix~\ref{app:proofs}.

\subsection{Step 1: A configuration-based fluid upper bound}
\label{subsec:dap_step_1}

Recall that $\mathfrak{A}_B$ denotes the collection of feasible
configurations defined in Section~\ref{subsec:dap_setting}.
For $\boldsymbol{A}=(A_1,\ldots,A_K)\in\mathfrak{A}_B$,
its common catalog is
$I_{\boldsymbol{A}}=\bigcup_{k=1}^K A_k$, with
$|I_{\boldsymbol{A}}|\leq B$.

Let $\tau_k$ denote the expected number of type-$k$ arrivals.
Thus, $\tau_k=\sum_{t=1}^T\lambda_{kt}$ when $T$ is
deterministic, and $\tau_k=\tau\lambda_k$ when $T$ is random
and customer types are independent and identically distributed.
For every configuration $\boldsymbol{A}$, define
\[
\mu_i^{\boldsymbol{A}}
=
\sum_{k=1}^K\tau_k\pi^k(i,A_k),
\qquad
D_{\boldsymbol{A}}
=
\sum_{i\in\mathcal{N}}\mu_i^{\boldsymbol{A}},
\qquad
V_{\boldsymbol{A}}
=
\sum_{i\in\mathcal{N}}r_i\mu_i^{\boldsymbol{A}}.
\]
These quantities describe the expected product-level demand,
total demand, and revenue when each type-$k$ customer is offered
$A_k$ without inventory restrictions.

\paragraph{The fluid relaxation.}
Introduce a nonnegative weight $f_{\boldsymbol{A}}$ for each
configuration and consider
\begin{equation}
\label{eqn:dap_configuration_lp}
\tag{FLUID-DAP}
\begin{aligned}
U_{\mathrm{conf}}
=
\max_{\boldsymbol{f}\geq\boldsymbol{0}}\quad
&\sum_{\boldsymbol{A}\in\mathfrak{A}_B}
f_{\boldsymbol{A}}V_{\boldsymbol{A}}\\
\text{s.t.}\quad
&\sum_{\boldsymbol{A}\in\mathfrak{A}_B}
f_{\boldsymbol{A}}D_{\boldsymbol{A}}\leq B,\\
&\sum_{\boldsymbol{A}\in\mathfrak{A}_B}
f_{\boldsymbol{A}}\leq1.
\end{aligned}
\end{equation}
The formulation has the same two-constraint structure as
\eqref{eqn:fluid_lp}. Its columns, however, specify a complete
personalized assortment plan rather than a single assortment.

\begin{lemma}[Configuration fluid upper bound]
\label{lem:dap_configuration_upper_bound}
The optimal value of~\eqref{eqn:dap_configuration_lp} satisfies
$U_{\mathrm{conf}}\geq\operatorname{OPT}_{\mathrm{DAP}}$.
\end{lemma}

To establish the lemma, we fix an initial inventory and an
adaptive policy, and record the distribution of assortments
offered to each customer type. These distributions can be
combined into a distribution over configurations supported
on the initial product catalog. The resulting fluid solution
preserves the policy's expected sales and revenue. In particular,
this construction does not partition physical inventory
across customer types.

\paragraph{Approximately solving the fluid relaxation.}
For $\lambda\geq0$, define
\[
H(\lambda)
=
\max_{\boldsymbol{A}\in\mathfrak{A}_B}
\{V_{\boldsymbol{A}}-\lambda D_{\boldsymbol{A}}\}.
\]
Using the definitions above,
\[
H(\lambda)
=
\max_{\boldsymbol{A}\in\mathfrak{A}_B}
\sum_{k=1}^K\tau_k
\sum_{i\in A_k}(r_i-\lambda)\pi^k(i,A_k)
=
\operatorname{JAC}_B
(\boldsymbol{r}-\lambda\boldsymbol{1}).
\]
Thus, the assumed JAC oracle gives an
$\alpha_{\mathrm{JAC}}$-approximation to the Lagrangian
subproblem for every capacity price $\lambda$.

\begin{lemma}[Approximate fluid solution]
\label{lem:dap_approximate_fluid}
For every $\varepsilon\in(0,1)$, a feasible solution to
\eqref{eqn:dap_configuration_lp} with objective value
$F\geq\alpha_{\mathrm{JAC}}(1-\varepsilon)U_{\mathrm{conf}}$
can be computed in time polynomial in the input encoding
and $\log(1/\varepsilon)$.
The returned solution is supported on at most two configurations.
\end{lemma}

The proof uses a bisection search over the capacity price.
We retain one configuration whose expected demand exceeds
$B$ and another whose expected demand is at most $B$.
Mixing these configurations to meet the capacity constraint
gives the desired approximation. This argument does not
require the approximate oracle's output to vary monotonically
with the capacity price.

Fix the solution $\boldsymbol{f}$ returned by
Lemma~\ref{lem:dap_approximate_fluid}, and let
$\mathfrak{A}^{+}
=\{\boldsymbol{A}:f_{\boldsymbol{A}}>0,\,
D_{\boldsymbol{A}}>0\}$.
Configurations with zero demand also have zero revenue and
can be discarded without changing $F$ or renormalizing the
remaining weights. If $F=0$, then
$\operatorname{OPT}_{\mathrm{DAP}}=0$, and the result is
immediate. Henceforth, assume $F>0$.

\subsection{Step 2: Constructing fractional inventory}
\label{subsec:dap_step_2}

For each $\boldsymbol{A}\in\mathfrak{A}^{+}$, we construct
a fractional inventory vector supported on
$I_{\boldsymbol{A}}$ and a corresponding revenue certificate.
The inventory is shared across customer types: its allocation
depends on the aggregate means
$\mu_i^{\boldsymbol{A}}$, rather than separate inventories
reserved for individual types.

\paragraph{The inventory construction.}
Let $d_{\boldsymbol{A}}=D_{\boldsymbol{A}}/B$.
For deterministic or Poisson arrivals, use the construction
from Section~\ref{subsec:step_2}, setting
$\beta_{\boldsymbol{A}}=1/(C+d_{\boldsymbol{A}})$ and
$\eta_{\boldsymbol{A}}=b(\beta_{\boldsymbol{A}})$, where
$C=1.0574$ and $b$ is defined
in~\eqref{eqn:inventory_slope}.

For IFR arrivals, use the construction from
Section~\ref{subsec:ifr_step_2}, setting
$\beta_{\boldsymbol{A}}
=1/(C_0+C_1d_{\boldsymbol{A}})$ and
$\eta_{\boldsymbol{A}}
=(1-\beta_{\boldsymbol{A}})/d_{\boldsymbol{A}}$,
where $C_0=1.2148$ and $C_1=1.0955$.

In either case, construct
\begin{equation}
\label{eqn:dap_fractional_inventory}
z_i^{\boldsymbol{A}}
=
\begin{cases}
\beta_{\boldsymbol{A}}
+
\eta_{\boldsymbol{A}}\mu_i^{\boldsymbol{A}},
&i\in I_{\boldsymbol{A}},\\
0,
&i\notin I_{\boldsymbol{A}}.
\end{cases}
\end{equation}
The feasibility calculations from the SA analysis imply
$\beta_{\boldsymbol{A}}
+\eta_{\boldsymbol{A}}d_{\boldsymbol{A}}\leq1$.
Consequently,
\[
\begin{aligned}
\sum_i z_i^{\boldsymbol{A}}
&=
\beta_{\boldsymbol{A}}|I_{\boldsymbol{A}}|
+
\eta_{\boldsymbol{A}}D_{\boldsymbol{A}}\\
&\leq
B\left(
\beta_{\boldsymbol{A}}
+
\eta_{\boldsymbol{A}}d_{\boldsymbol{A}}
\right)
\leq B.
\end{aligned}
\]
The common-catalog restriction supplies precisely the bound
$|I_{\boldsymbol{A}}|\leq B$ needed here.

\paragraph{An expected-revenue lower bound.}
For a fixed configuration $\boldsymbol{A}$, let
$\sigma^{\boldsymbol{A}}$ denote the policy that offers
$A_k\cap\{i:c_{i,t}>0\}$ whenever a type-$k$ customer arrives.
This is a type-specific show-all policy: every available
product in the type's predetermined assortment is offered.

For a nonnegative vector $\boldsymbol{z}$ supported on
$I_{\boldsymbol{A}}$, define
\[
\mathcal{R}_{\boldsymbol{A}}(\boldsymbol{z})
=
\begin{cases}
\displaystyle
\sum_{i\in I_{\boldsymbol{A}}}
r_i g_{\mu_i^{\boldsymbol{A}}}(z_i),
&\text{for deterministic or Poisson arrivals},\\[3mm]
\displaystyle
\sum_{i\in I_{\boldsymbol{A}}}
r_i g_{\mu_i^{\boldsymbol{A}}}^{\mathrm{G}}(z_i),
&\text{for IFR arrivals}.
\end{cases}
\]
Here, $g_\mu$ and $g_\mu^{\mathrm{G}}$ are the Poisson and
geometric truncated-demand functions used in the preceding
sections.

\begin{lemma}[Personalized revenue lower bound]
\label{lem:dap_implementation}
For every configuration $\boldsymbol{A}\in\mathfrak{A}_B$
and integral inventory vector $\boldsymbol{c}$ supported on
$I_{\boldsymbol{A}}$,
\[
\mathrm{Rev}^{\mathrm{DAP}}
(\boldsymbol{c},\sigma^{\boldsymbol{A}})
\geq
\mathcal{R}_{\boldsymbol{A}}(\boldsymbol{c}).
\]
\end{lemma}

The proof applies substitutability separately to each customer
type. While a planned product remains available, removing
stocked-out products from that type's assortment cannot decrease
its purchase probability. Hypothetical demand is
Poisson-binomial for deterministic arrivals, exactly Poisson
under Poisson arrivals with independent and identically
distributed types, and dominated in convex order by geometric
demand under IFR arrivals. The demand comparisons from the
SA analysis therefore remain applicable after aggregating
across types.

\paragraph{A configuration-level revenue certificate.}
For deterministic or Poisson arrivals,
Lemma~\ref{lem:poisson_inventory_bound} gives
$g_{\mu_i^{\boldsymbol{A}}}
(z_i^{\boldsymbol{A}})
\geq\beta_{\boldsymbol{A}}\mu_i^{\boldsymbol{A}}$.

For IFR arrivals, the admissibility condition
\eqref{eqn:ifr_admissible_constants} implies
$\eta_{\boldsymbol{A}}
\geq b_{\mathrm{G}}(\beta_{\boldsymbol{A}})$.
Monotonicity and Lemma~\ref{lem:ifr_inventory_bound} therefore
give the same comparison with $g^{\mathrm{G}}$.
Thus, in either setting,
\begin{equation}
\label{eqn:dap_configuration_certificate}
\mathcal{R}_{\boldsymbol{A}}
(\boldsymbol{z}^{\boldsymbol{A}})
\geq
\beta_{\boldsymbol{A}}V_{\boldsymbol{A}}
=:L_{\boldsymbol{A}}.
\end{equation}
In particular,
\[
L_{\boldsymbol{A}}
=
\begin{cases}
\displaystyle
\frac{V_{\boldsymbol{A}}}{C+d_{\boldsymbol{A}}},
&\text{for deterministic or Poisson arrivals},\\[3mm]
\displaystyle
\frac{V_{\boldsymbol{A}}}{C_0+C_1d_{\boldsymbol{A}}},
&\text{for IFR arrivals}.
\end{cases}
\]

\paragraph{Selecting the best candidate.}
Choose
$\widehat{\boldsymbol{A}}
\in\arg\max_{\boldsymbol{A}\in\mathfrak{A}^{+}}
L_{\boldsymbol{A}}$,
and retain
$\boldsymbol{z}^{\widehat{\boldsymbol{A}}}$.
To combine the two arrival settings, write
$(\kappa_0,\kappa_1)=(C,1)$ for deterministic or Poisson
arrivals and $(\kappa_0,\kappa_1)=(C_0,C_1)$ for IFR arrivals.
The two fluid constraints imply
\begin{equation}
\label{eqn:dap_combine_certificates}
\begin{aligned}
F
&=
\sum_{\boldsymbol{A}\in\mathfrak{A}^{+}}
f_{\boldsymbol{A}}V_{\boldsymbol{A}}\\
&=
\sum_{\boldsymbol{A}\in\mathfrak{A}^{+}}
f_{\boldsymbol{A}}
\left(
\kappa_0+\kappa_1\frac{D_{\boldsymbol{A}}}{B}
\right)L_{\boldsymbol{A}}\\
&\leq
L_{\widehat{\boldsymbol{A}}}
\left(
\kappa_0\sum_{\boldsymbol{A}\in\mathfrak{A}^{+}}
f_{\boldsymbol{A}}
+
\frac{\kappa_1}{B}
\sum_{\boldsymbol{A}\in\mathfrak{A}^{+}}
f_{\boldsymbol{A}}D_{\boldsymbol{A}}
\right)\\
&\leq
(\kappa_0+\kappa_1)L_{\widehat{\boldsymbol{A}}}.
\end{aligned}
\end{equation}
Therefore, the selected fractional inventory satisfies
\[
\mathcal{R}_{\widehat{\boldsymbol{A}}}
(\boldsymbol{z}^{\widehat{\boldsymbol{A}}})
\geq
L_{\widehat{\boldsymbol{A}}}
\geq
\frac{F}{\kappa_0+\kappa_1}.
\]

\subsection{Step 3: Randomized rounding and implementation}
\label{subsec:dap_step_3}

Apply the dependent-rounding procedure from
Section~\ref{subsec:step_3} to
$\boldsymbol{z}^{\widehat{\boldsymbol{A}}}$, independently
of future customer arrivals and preferences. Let
$\widehat{\boldsymbol{c}}(\omega)$ denote the resulting
integral inventory vector. The same rounding properties
ensure that it is supported on
$I_{\widehat{\boldsymbol{A}}}$ and belongs to
$\mathcal{C}_B$ almost surely.

Both $g_\mu$ and $g_\mu^{\mathrm{G}}$ are affine between
consecutive integers. Hence, the argument in
Lemma~\ref{lem:rounding_preserves_surrogate} applies to the
configuration surrogate and gives
\[
\mathbb{E}_{\omega}
\left[
\mathcal{R}_{\widehat{\boldsymbol{A}}}
(\widehat{\boldsymbol{c}}(\omega))
\right]
=
\mathcal{R}_{\widehat{\boldsymbol{A}}}
(\boldsymbol{z}^{\widehat{\boldsymbol{A}}}).
\]

Set $\widehat\sigma=\sigma^{\widehat{\boldsymbol{A}}}$.
The algorithm implements this policy throughout the selling
horizon: when type $k$ arrives, it offers
$\widehat A_k$ intersected with the current in-stock set.
The fluid mixture is used only to identify the selected
configuration; it is not sampled anew during the selling
horizon.

\paragraph{Completing the approximation guarantee.}
Using Lemma~\ref{lem:dap_implementation},
\eqref{eqn:dap_configuration_certificate},
\eqref{eqn:dap_combine_certificates}, and
Lemma~\ref{lem:dap_approximate_fluid}, we obtain
\[
\begin{aligned}
\mathbb{E}_{\omega}
\left[
\mathrm{Rev}^{\mathrm{DAP}}
(\widehat{\boldsymbol{c}}(\omega),\widehat\sigma)
\right]
&\geq
\mathbb{E}_{\omega}
\left[
\mathcal{R}_{\widehat{\boldsymbol{A}}}
(\widehat{\boldsymbol{c}}(\omega))
\right]\\
&=
\mathcal{R}_{\widehat{\boldsymbol{A}}}
(\boldsymbol{z}^{\widehat{\boldsymbol{A}}})\\
&\geq
L_{\widehat{\boldsymbol{A}}}\\
&\geq
\frac{F}{\kappa_0+\kappa_1}\\
&\geq
\frac{\alpha_{\mathrm{JAC}}(1-\varepsilon)}
{\kappa_0+\kappa_1}
U_{\mathrm{conf}}\\
&\geq
\frac{\alpha_{\mathrm{JAC}}(1-\varepsilon)}
{\kappa_0+\kappa_1}
\operatorname{OPT}_{\mathrm{DAP}}.
\end{aligned}
\]
The revenue function already averages over customer arrivals,
types, and purchase decisions; the outer expectation concerns
only the initial rounding randomness.

Taking $\varepsilon=10^{-4}$ gives
\[
\frac{1-\varepsilon}{1+C}
=
\frac{9999}{20574}
>
0.486,
\qquad
\frac{1-\varepsilon}{C_0+C_1}
=
\frac{9999}{23103}
>
0.4328.
\]
Consequently, the algorithm achieves the stated guarantees of
$0.486\alpha_{\mathrm{JAC}}$ for deterministic or Poisson
arrivals and $0.4328\alpha_{\mathrm{JAC}}$ for IFR arrivals.

\paragraph{Computational complexity and policy structure.}
Lemma~\ref{lem:dap_approximate_fluid} computes a fluid solution
with at most two supporting configurations using polynomially
many calls to the JAC oracle. The candidate inventories and
certificates are evaluated using rational arithmetic, and
rounding has the same expected polynomial running time as in
Section~\ref{subsec:step_3}. During the selling horizon, the
policy merely identifies the arriving type and removes
unavailable products from its predetermined assortment.

In particular, the policy never strategically withholds an
available product from its planned assortment, introduces a
previously excluded product following a stockout, or resolves
an assortment optimization problem during the horizon.
Nevertheless, its performance is compared with the optimal
adaptive DAP policy. This completes the proof of
Theorem~\ref{thm:dap_approximation}.

\makeatletter
\newcommand*\mysize{%
\@setfontsize\mysize{10.0}{11.0}%
}
\makeatother

\renewcommand{\bibfont}{\mysize}
{\setlength{\bibsep}{3pt}
\bibliographystyle{plainnat}
\bibliography{bib_size-sub.bib}
}

\ECSwitch


\ECHead{Electronic Companion}


\begin{appendices}

\section{Proof from Section~\ref{sec:proof_main_thm}} \label{app:proofs}

\subsection{Proof of Lemma~\ref{lem:revenue_lower_bound}}

Consider the hypothetical demand generated by offering $S$ to every
customer, irrespective of inventory availability. Let $Y_i^S$
denote the number of customers who select product $i$ in this
hypothetical process. Conditional on $T$, its marginal distribution
is $\operatorname{Binomial}(T,\pi(i,S))$.

Under the actual show-all process, every offered assortment is a
subset of $S$. Whenever product $i$ remains available,
substitutability implies that its purchase probability is at least
$\pi(i,S)$. A productwise coupling therefore gives
\begin{equation}
\label{eqn:fixed_assortment_lower_bound}
\mathrm{Rev}(\boldsymbol{c})
\geq
\sum_{i\in S}r_i\expar{\min\{c_i,Y_i^S\}}.
\end{equation}
This is the fixed-assortment revenue comparison discussed in
Remark~2 and the proof of Proposition~2 of~\cite{fu2026joint}.

When $T$ is Poisson distributed, thinning gives
$Y_i^S\sim\operatorname{Poisson}(\mu_i^S)$, and the result follows
directly. When $T$ is deterministic, $Y_i^S$ is binomial with mean
$\mu_i^S$. A binomial random variable is dominated in convex order
by a Poisson random variable with the same mean. Since
$x\mapsto\min\{c_i,x\}$ is concave, it follows that
\[
\expar{\min\{c_i,Y_i^S\}}
\geq
\expar{\min\{c_i,X_{\mu_i^S}\}}
=
g_{\mu_i^S}(c_i).
\]

For completeness, the convex-order comparison can be established
by replacing the independent Bernoulli summands one at a time by
independent Poisson variables with the same means. For any concave
function $\varphi$ and nonnegative integer $k$,
$\varphi(k)\leq\varphi(0)+k(\varphi(1)-\varphi(0))$.
Thus, among nonnegative integer-valued random variables of mean
$p\in[0,1]$, a Bernoulli random variable maximizes the expectation
of $\varphi$. Conditioning on the other summands and applying this
observation successively proves the required comparison.
Combining it with~\eqref{eqn:fixed_assortment_lower_bound} completes
the proof.

\subsection{Proof of Lemma~\ref{lem:poisson_inventory_bound}}

The case $\mu=0$ is immediate, so suppose $\mu>0$. We first consider
$\beta\in[1/2,1)$. Let $X\sim\operatorname{Poisson}(\mu)$, and write
\[
p=\Pr(X>0)=1-e^{-\mu},
\qquad
\vartheta=\Pr(X\geq2)=1-(1+\mu)e^{-\mu}.
\]
Using $2\min\{X,q\}=X+q-|X-q|$ on the event $X>0$, followed by
Cauchy--Schwarz, gives
\begin{align}
2g_\mu(q)
&=
\mu+pq-
\expar{|X-q|\mathbf{1}\{X>0\}}
\nonumber\\
&\geq
\mu+pq-
\sqrt{p\,\expar{(X-q)^2\mathbf{1}\{X>0\}}}
\nonumber\\
&=
\mu+pq-
\sqrt{p(\mu+\mu^2-2q\mu+pq^2)}
\nonumber\\
&=
\mu+pq-
\sqrt{(pq-\mu)^2+\mu\vartheta}.
\label{eqn:conditional_cs_bound}
\end{align}
Here, we use $\expar{X^2}=\mu+\mu^2$ and
$p(1+\mu)-\mu=\vartheta$. Restricting the application of
Cauchy--Schwarz to $X>0$ treats the zero-demand event exactly.

Set $a=1/(4(1-\beta))$ and
$
q_0=\frac{\beta\mu+a\vartheta}{p}.
$
Because $4a\cdot (1-\beta)=1$, we have
$
(pq_0-\mu)^2+\mu\vartheta
=
\bigl((1-\beta)\mu+a\vartheta\bigr)^2.
$
Substituting into~\eqref{eqn:conditional_cs_bound} yields
\begin{equation}
\label{eqn:exact_scalar_inventory}
g_\mu(q_0)
\geq
\frac{
\mu+\beta\mu+a\vartheta
-
((1-\beta)\mu+a\vartheta)
}{2}
=
\beta\mu.
\end{equation}

It remains to upper-bound $q_0$ by the inventory prescribed in the
lemma. Let $v=\mu/(e^\mu-1)$. The identities
$\mu/p=\mu+v$ and $\vartheta/p=1-v$ imply
\[
q_0
=
\beta(\mu+v)+a(1-v)
=
\beta+\beta\mu+(a-\beta)(1-v).
\]
Moreover,
\[
a-\beta
=
\frac{(2\beta-1)^2}{4(1-\beta)}
\geq0,
\qquad
1-\frac{\mu}{e^\mu-1}\leq\frac{\mu}{2}.
\]
To verify the second inequality, define
$H(\mu)=(\mu-2)e^\mu+\mu+2$. The inequality is equivalent to
$H(\mu)\geq0$, which follows from
$H(0)=H'(0)=0$ and $H''(\mu)=\mu e^\mu\geq0$.
Consequently,
\[
q_0
\leq
\beta+\beta\mu+\frac{a-\beta}{2}\mu
=
\beta+
\left(
\frac{\beta}{2}+\frac{1}{8(1-\beta)}
\right)\mu
=
\beta+b(\beta)\mu.
\]
Since $g_\mu$ is nondecreasing,
\eqref{eqn:exact_scalar_inventory} proves the claim for
$\beta\geq1/2$.

Finally, suppose $\beta\leq1/2$. The result at $\beta=1/2$ gives
$g_\mu((1+\mu)/2)\geq\mu/2$.
The function $g_\mu$ is concave and satisfies $g_\mu(0)=0$, so
\[
g_\mu\bigl(\beta(1+\mu)\bigr)
=
g_\mu\left(2\beta\frac{1+\mu}{2}\right)
\geq
2\beta g_\mu\left(\frac{1+\mu}{2}\right)
\geq
\beta\mu.
\]
Since $b(\beta)=\beta$ in this case, the proof is complete.

\subsection{Proof of Lemma~\ref{lem:rounding_preserves_surrogate}}

Let $\omega$ denote the rounding randomness, and write
$\widehat{\boldsymbol{c}}=\widehat{\boldsymbol{c}}(\omega)$.
Fix $i\in\widehat S$, and abbreviate
$\mu_i=\mu_i^{\widehat S}$.
Recall that
$g_{\mu_i}(q)=\mathbb{E}_{X_{\mu_i}}[\min\{q,X_{\mu_i}\}]$,
where $X_{\mu_i}$ is an auxiliary Poisson random variable with
mean $\mu_i$, independent of $\omega$. Thus, $g_{\mu_i}(q)$
already averages over auxiliary demand and is deterministic
for a fixed inventory level $q$.

Since $X_{\mu_i}$ is integer-valued, for every integer $k\geq0$
and $\theta\in[0,1]$,
\[
\min\{k+\theta,X_{\mu_i}\}
=
(1-\theta)\min\{k,X_{\mu_i}\}
+
\theta\min\{k+1,X_{\mu_i}\}.
\]
Taking expectations with respect to $X_{\mu_i}$ gives
\begin{equation}
\label{eqn:truncated_demand_interpolation}
g_{\mu_i}(k+\theta)
=
(1-\theta)g_{\mu_i}(k)
+
\theta g_{\mu_i}(k+1).
\end{equation}

Now write $z_i=k_i+a_i$, where $k_i=\lfloor z_i\rfloor$
and $a_i\in[0,1)$. The marginal rounding probabilities satisfy
$\Pr_{\omega}\{\widehat c_i=k_i+1\}=a_i$ and
$\Pr_{\omega}\{\widehat c_i=k_i\}=1-a_i$.
Consequently, taking expectations only over the rounding
randomness,
\[
\begin{aligned}
\mathbb{E}_{\omega}
\left[g_{\mu_i}(\widehat c_i(\omega))\right]
&=
(1-a_i)g_{\mu_i}(k_i)
+
a_i g_{\mu_i}(k_i+1)\\
&=
g_{\mu_i}(k_i+a_i)
=
g_{\mu_i}(z_i),
\end{aligned}
\]
where the second equality follows
from~\eqref{eqn:truncated_demand_interpolation}.
Multiplying by $r_i$ and summing over products yields
\[
\begin{aligned}
\mathbb{E}_{\omega}
\left[
\mathcal{R}(\widehat{\boldsymbol{c}}(\omega))
\right]
&=
\sum_{i\in\widehat S}
r_i\mathbb{E}_{\omega}
\left[g_{\mu_i}(\widehat c_i(\omega))\right]\\
&=
\sum_{i\in\widehat S}r_i g_{\mu_i}(z_i)
=
\mathcal{R}(\boldsymbol{z}).
\end{aligned}
\]
This argument uses only the marginal rounding probabilities;
the coordinates of $\widehat{\boldsymbol{c}}$ need not be
independent. The expectation over auxiliary demand is already
incorporated into $\mathcal{R}$, so the outer expectation above
concerns only the rounding randomness.

\subsection{Proof Proposition~\ref{prop:fluid_gap}}

Fix an integer $m\geq2$. Consider an instance with two products, inventory capacity $B=1$, and a deterministic number $T=m$ of customer arrivals. Customer choice follows an MNL model with outside-option weight $w_0=1$ and product weights
\[
w_1=\frac{1}{m^2-1},
\qquad
w_2=m.
\]
The product revenues are
\[
r_1=m,
\qquad
r_2=1.
\]
The singleton purchase probabilities are therefore
\[
\pi(1,\{1\})
=
\frac{w_1}{1+w_1}
=
\frac{1}{m^2},
\qquad
\pi(2,\{2\})
=
\frac{w_2}{1+w_2}
=
\frac{m}{m+1}.
\]

We first characterize the optimal inventory decision. Since $B=1$, the DM can stock at most one unit of one product. If product $1$ is stocked, it generates revenue $m$ if at least one of the $m$ customers purchases it. Hence,
\[
\mathrm{Rev}(\boldsymbol{e}_1)
=
m\left[
1-\left(1-\frac{1}{m^2}\right)^m
\right]
\leq 1,
\]
where the inequality follows from
$1-(1-x)^m\leq mx$. Moreover,
\[
\mathrm{Rev}(\boldsymbol{e}_1)
\longrightarrow1.
\]
If product $2$ is stocked, then
\[
\mathrm{Rev}(\boldsymbol{e}_2)
=
1-
\left(\frac{1}{m+1}\right)^m
\longrightarrow1.
\]
Consequently,
\begin{equation}
\label{eqn:gap_stochastic_opt}
\operatorname{OPT}_{\mathrm{SA}}
=
\max\{
\mathrm{Rev}(\boldsymbol{e}_1),
\mathrm{Rev}(\boldsymbol{e}_2)
\}
\longrightarrow1.
\end{equation}
Since only one product can be stocked, dynamically varying the offered assortment cannot improve revenue. Thus, the same conclusion holds even if the DM is allowed to dynamically select assortments.

We next consider the fluid relaxation. Since $B=1$, the only nonempty assortments that can receive positive weight are $\{1\}$ and $\{2\}$. Their expected demands and revenues over the full horizon are
\[
D_{\{1\}}
=
m\pi(1,\{1\})
=
\frac1m,
\qquad
V_{\{1\}}
=
mr_1\pi(1,\{1\})
=
1,
\]
and
\[
D_{\{2\}}
=
m\pi(2,\{2\})
=
\frac{m^2}{m+1},
\qquad
V_{\{2\}}
=
\frac{m^2}{m+1}
=
D_{\{2\}}.
\]

For notational convenience, write $D_1=D_{\{1\}}$ and $D_2=D_{\{2\}}$. Consider the fluid solution
\[
f_{\{2\}}
=
\frac{1-D_1}{D_2-D_1},
\qquad
f_{\{1\}}
=
1-f_{\{2\}}.
\]
Both coefficients are nonnegative for $m\geq2$, and by construction
\[
f_{\{1\}}+f_{\{2\}}=1,
\qquad
f_{\{1\}}D_1+f_{\{2\}}D_2=1.
\]
Hence, this is a feasible solution to~\eqref{eqn:fluid_lp}. Since
$D_1\rightarrow0$, $D_2\rightarrow\infty$, and
$f_{\{2\}}=(1-D_1)/(D_2-D_1)\rightarrow0$, we have
\[
f_{\{2\}}D_2\longrightarrow1
\qquad\text{and}\qquad
f_{\{1\}}\longrightarrow1.
\]
Its objective value therefore satisfies
\[
f_{\{1\}}V_{\{1\}}
+
f_{\{2\}}V_{\{2\}}
=
f_{\{1\}}
+
f_{\{2\}}D_2
\longrightarrow2.
\]
Thus,
\[
\liminf_{m\rightarrow\infty}
\operatorname{OPT}_{\mathrm{FL}}
\geq2.
\]

Conversely, for any feasible fluid solution $(f_{\{1\}},f_{\{2\}})$,
\[
\begin{aligned}
f_{\{1\}}V_{\{1\}}
+
f_{\{2\}}V_{\{2\}}
&=
f_{\{1\}}
+
f_{\{2\}}D_2\\
&=
f_{\{1\}}D_1
+
f_{\{2\}}D_2
+
f_{\{1\}}(1-D_1)\\
&\leq
1+f_{\{1\}}(1-D_1)
\leq2.
\end{aligned}
\]
Therefore,
\begin{equation}
\label{eqn:gap_fluid_opt}
\operatorname{OPT}_{\mathrm{FL}}
\longrightarrow2.
\end{equation}
Combining~\eqref{eqn:gap_stochastic_opt} and~\eqref{eqn:gap_fluid_opt} gives
\[
\frac{\operatorname{OPT}_{\mathrm{SA}}}
{\operatorname{OPT}_{\mathrm{FL}}}
\longrightarrow
\frac12,
\]
as required.

\section{Proofs from Section~\ref{sec:proof_ifr}}

\subsection{Proof of Lemma~\ref{lem:ifr_revenue_lower_bound}}

Fix an assortment $S$ and an integral inventory vector
$\boldsymbol{c}$ supported on $S$. Let $X_i^S$ denote
hypothetical demand for product $i$ when $S$ is offered to
every customer without inventory restrictions.
As in the proof of Lemma~\ref{lem:revenue_lower_bound},
substitutability gives
\[
\mathrm{Rev}(\boldsymbol{c})
\geq
\sum_{i\in S}
r_i\mathbb{E}[\min\{c_i,X_i^S\}].
\]
It remains to compare $X_i^S$ with a geometric random variable.
Conditional on $T$,
\[
X_i^S\mid T
\sim
\operatorname{Binomial}(T,\pi(i,S)).
\]

We first show that any nonnegative integer-valued IFR random
variable $T$ with mean $\tau$ satisfies
$T\leq_{\mathrm{cx}}G_\tau$.
Let $s_k=\Pr(T\geq k)$.
The IFR condition implies
$s_{k+j}\leq s_ks_j$ for all $k,j\geq0$.
Indeed, wherever $s_k>0$, this follows by multiplying the
nonincreasing continuation probabilities
$s_{\ell+1}/s_\ell$; if $s_k=0$, the inequality is immediate.

Define $H(k)=\mathbb{E}[(T-k)^+]$. Then
\[
H(k)
=
\sum_{j=1}^{\infty}s_{k+j}
\leq
s_k\sum_{j=1}^{\infty}s_j
=
\tau s_k.
\]
For $k\geq1$, we also have $H(k-1)=s_k+H(k)$, and hence
\[
H(k)
\leq
\frac{\tau}{1+\tau}H(k-1).
\]
Since $H(0)=\tau$, induction gives
\[
\mathbb{E}[(T-k)^+]
\leq
\tau
\left(\frac{\tau}{1+\tau}\right)^k
=
\mathbb{E}[(G_\tau-k)^+]
\]
for all integers $k\geq0$.
The comparison extends to all real thresholds by linear
interpolation, since both random variables are integer-valued.
As they have the same mean, the stop-loss characterization
of convex order yields
$T\leq_{\mathrm{cx}}G_\tau$.

Next, let $B_1,B_2,\ldots$ be independent
$\operatorname{Bernoulli}(p)$ random variables, independent of
the number of summands. For any convex function $\varphi$,
define
$
F_\varphi(n)
=
\mathbb{E}[\varphi(\sum_{j=1}^nB_j)].
$
Writing $Y_n=\sum_{j=1}^nB_j$, we obtain
\[
\begin{aligned}
&F_\varphi(n+2)-2F_\varphi(n+1)+F_\varphi(n)\\
&\quad=
p^2\mathbb{E}\left[
\varphi(Y_n+2)-2\varphi(Y_n+1)+\varphi(Y_n)
\right]
\geq0.
\end{aligned}
\]
Thus, $F_\varphi$ is a convex sequence, so independent
Bernoulli thinning preserves convex order:
\[
\sum_{j=1}^{T}B_j
\leq_{\mathrm{cx}}
\sum_{j=1}^{G_\tau}B_j.
\]
Moreover, thinning a geometric random variable with mean
$\tau$ by probability $p$ gives a geometric random variable
with mean $\tau p$, as can be verified from its probability
generating function:
\[
\mathbb{E}[(1-p+pz)^{G_\tau}]
=
\frac{1}{1+\tau p(1-z)}.
\]

Taking $p=\pi(i,S)$, we conclude that
$X_i^S\leq_{\mathrm{cx}}G_{\mu_i^S}$.
Since $x\mapsto\min\{c_i,x\}$ is concave,
\[
\mathbb{E}[\min\{c_i,X_i^S\}]
\geq
\mathbb{E}[\min\{c_i,G_{\mu_i^S}\}]
=
g_{\mu_i^S}^{\mathrm{G}}(c_i).
\]
Multiplying by $r_i$ and summing over products proves
the lemma.

\subsection{Proof of Lemma~\ref{lem:ifr_inventory_bound}}

The case $\mu=0$ is immediate, so suppose $\mu>0$.
Let $a=\mu/(1+\mu)$. Since
$\Pr(G_\mu\geq k)=a^k$, for every integer $k\geq0$ and
$\theta\in[0,1]$,
\begin{equation}
\label{eqn:ifr_geometric_truncation}
g_\mu^{\mathrm{G}}(k+\theta)
=
\mu(1-a^k)+\theta a^{k+1}
=
\mu\left[
1-a^k\left(1-\frac{\theta}{1+\mu}\right)
\right].
\end{equation}

\paragraph{Case 1: $\beta\leq7/10$.}
Let $\ell=-\log(1-\beta)$, and define
\[
s_\beta(\mu)
=
\min\{q\geq0:
g_\mu^{\mathrm{G}}(q)\geq\beta\mu\}.
\]
We show that $s_\beta(\mu)\leq\beta+\ell\mu$.

If $s_\beta(\mu)\leq1$, then
$g_\mu^{\mathrm{G}}(q)=q\mu/(1+\mu)$
on this interval, so
\[
s_\beta(\mu)
=
\beta(1+\mu)
\leq
\beta+\ell\mu,
\]
where we use $\ell\geq\beta$.

Otherwise, write $s_\beta(\mu)=k+\theta$, with
$k\geq1$ an integer and $\theta\in[0,1)$.
Setting
$g_\mu^{\mathrm{G}}(k+\theta)=\beta\mu$
in~\eqref{eqn:ifr_geometric_truncation}, and introducing
$x=1/a>1$, gives
\begin{equation}
\label{eqn:ifr_inverse_slope}
\frac{s_\beta(\mu)-\beta}{\mu}
=
(k+1-\beta)x-(k-\beta)
-(1-\beta)x^{k+1}.
\end{equation}
Maximizing the right-hand side over $x\geq1$ yields
\[
M_k(\beta)
=
k\left(1-\frac{\beta}{k+1}\right)
\exp\left(
\frac{\ell+\log(1-\beta/(k+1))}{k}
\right)
-k+\beta.
\]
We claim that $M_k(\beta)\leq\ell$ for all $k\geq1$.

For $k=1$, direct simplification gives
\[
M_1(\beta)
=
\beta+\frac{\beta^2}{4(1-\beta)}.
\]
Moreover,
\[
\frac{d}{d\beta}
\left[
-\log(1-\beta)-\beta
-\frac{\beta^2}{4(1-\beta)}
\right]
=
\frac{\beta(2-3\beta)}
{4(1-\beta)^2}.
\]
The expression in brackets increases up to $\beta=2/3$
and then decreases. Its value at zero is zero, while its
value at $7/10$ is
$\log(10/3)-133/120>0$.
Therefore, $M_1(\beta)\leq\ell$ for
$\beta\in(0,7/10]$.

Now suppose $k\geq2$, and let $\delta=\ell-\beta$.
Since $\ell/\beta$ is increasing in $\beta$ and
$\log(10/3)<49/40$, we have
$0\leq\delta\leq3\beta/4$ throughout the interval.
Using $\log(1-u)\leq-u-u^2/2$ for $u\in[0,1)$,
\[
(k+1)\log\left(1-\frac{\beta}{k+1}\right)
+\ell
\leq
\delta-\frac{\beta^2}{2(k+1)}.
\]
Since $\delta^2\leq9\beta^2/16$ and $k\geq2$,
we have
$\delta^2/k\leq\beta^2/(k+1)$.
Consequently,
\[
\begin{aligned}
&(k+1)\log\left(1-\frac{\beta}{k+1}\right)
+\ell\\
&\qquad\leq
\delta-\frac{\delta^2}{2k}\\
&\qquad\leq
k\log\left(1+\frac{\delta}{k}\right),
\end{aligned}
\]
where the last inequality uses
$\log(1+v)\geq v-v^2/2$ for $v\geq0$.
Exponentiating gives
\[
\left(1-\frac{\beta}{k+1}\right)
\exp\left(
\frac{\ell+\log(1-\beta/(k+1))}{k}
\right)
\leq
1+\frac{\delta}{k}.
\]
Substituting into the expression for $M_k(\beta)$ yields
\[
M_k(\beta)
\leq
k\left(1+\frac{\delta}{k}\right)-k+\beta
=
\ell.
\]
Together with~\eqref{eqn:ifr_inverse_slope},
this proves
$s_\beta(\mu)\leq\beta+\ell\mu$, and hence
\[
g_\mu^{\mathrm{G}}
\bigl(\beta-\mu\log(1-\beta)\bigr)
\geq
\beta\mu.
\]

\paragraph{Case 2: $\beta>7/10$.}
Let $G=G_\mu$ and
$p=\Pr(G>0)=\mu/(1+\mu)$.
Since $\mathbb{E}[G^2]=\mu+2\mu^2$, applying
Cauchy--Schwarz on the event $G>0$ gives
\[
\begin{aligned}
2g_\mu^{\mathrm{G}}(q)
&=
\mu+pq
-\mathbb{E}[|G-q|\mathbf{1}\{G>0\}]\\
&\geq
\mu+pq-
\sqrt{p\,
\mathbb{E}[(G-q)^2\mathbf{1}\{G>0\}]}\\
&=
\mu+pq-
\sqrt{(pq-\mu)^2+\mu^2p}.
\end{aligned}
\]
Set $a_\beta=1/[4(1-\beta)]$ and
$q_0=\beta+(\beta+a_\beta)\mu$.
Then
\[
pq_0
=
\beta\mu+a_\beta\mu p
\]
and, using $4a_\beta(1-\beta)=1$,
\[
(pq_0-\mu)^2+\mu^2p
=
\mu^2(1-\beta+a_\beta p)^2.
\]
Substituting into the preceding inequality yields
\[
g_\mu^{\mathrm{G}}(q_0)
\geq
\frac{
\mu+\beta\mu+a_\beta\mu p
-\mu(1-\beta+a_\beta p)
}{2}
=
\beta\mu.
\]
This completes the proof.

\subsection{Proof of Lemma~\ref{lem:ifr_admissible_constants}}

We first verify the coefficient condition for
$C_0^\star,C_1^\star$.

\paragraph{The logarithmic branch.}
For $d>0$, let $\beta_0(d)\in(0,1)$ be the unique
solution to
\[
\beta_0(d)-d\log(1-\beta_0(d))=1,
\]
and define $H(d)=1/\beta_0(d)$.
Introducing $\ell=-\log(1-\beta_0(d))$
gives the parameterization
\[
d=\frac{e^{-\ell}}{\ell},
\qquad
H(d)=\frac{1}{1-e^{-\ell}},
\qquad \ell>0.
\]
Differentiating,
\[
H'(d)
=
\frac{\ell^2}
{(1+\ell)(1-e^{-\ell})^2}.
\]
The logarithmic derivative of the right-hand side with respect
to $\ell$ is
\[
\frac{2}{\ell}
-\frac{1}{1+\ell}
-\frac{2}{e^\ell-1}
\geq0.
\]
Indeed, the inequality is equivalent to
$(\ell+2)(e^\ell-1)\geq2\ell(\ell+1)$,
which follows from
$e^\ell-1\geq\ell+\ell^2/2$.
Since $d=e^{-\ell}/\ell$ decreases with $\ell$,
it follows that $H'(d)$ is nonincreasing in $d$.
Therefore, $H$ is concave.

At $d=1$, the defining equation gives
$\ell=e^{-\ell}$, so $\ell=w$ and
$\beta_0(1)=1-w$.
The tangent to $H$ at $d=1$ has slope
\[
H'(1)
=
\frac{w^2}{(1+w)(1-w)^2}
=
C_1^\star
\]
and intercept
$
H(1)-H'(1)
=
1/(1-w)-C_1^\star
=
C_0^\star.
$
By concavity,
\[
H(d)\leq C_0^\star+C_1^\star d,
\qquad d>0.
\]

For any $\beta\in(0,1)$, set
$d=(1-\beta)/[-\log(1-\beta)]$.
Then $\beta_0(d)=\beta$, so the tangent inequality gives
\[
\frac{1}{\beta}
-
C_1^\star
\frac{1-\beta}{-\log(1-\beta)}
\leq C_0^\star.
\]
This proves the required condition on the logarithmic branch.

\paragraph{The remaining branch.}
For $\beta\in[7/10,1)$, define
\[
J(\beta)
=
\frac{1}{\beta}
-
\frac{4C_1^\star(1-\beta)^2}
{1+4\beta(1-\beta)}.
\]
Writing $c=1-\beta\in(0,3/10]$, differentiation gives
\[
J'(\beta)
=
-\frac{1}{\beta^2}
+
\frac{8C_1^\star c(1+2c)}
{(1+4c-4c^2)^2}.
\]
Using $C_1^\star<11/10$, we have
\[
\begin{aligned}
&(1+4c-4c^2)^2
-8C_1^\star(1-c)^2c(1+2c)\\
&\qquad\geq
1-\frac45c+8c^2
-\frac{28}{5}c^3-\frac85c^4\\
&\qquad\geq
1-\frac45\left(\frac3{10}\right)
-\frac{28}{5}\left(\frac3{10}\right)^3
-\frac85\left(\frac3{10}\right)^4\\
&\qquad=
\frac{1862}{3125}>0.
\end{aligned}
\]
Hence $J'(\beta)<0$ on this interval, implying
\[
J(\beta)
\leq
J(7/10)
=
\frac{10}{7}-\frac9{46}C_1^\star
<
C_0^\star.
\]
Thus, the admissibility condition holds on both branches.

Finally, numerical bounds obtained from
$0.56714329<w<0.56714330$ give
\[
C_0^\star<1.2148,
\qquad
C_1^\star<1.0955.
\]
The enclosure for $w$ can be verified by evaluating $xe^x$
at the two rational endpoints and bounding the exponential
series by a finite Taylor expansion with a geometric remainder.
Increasing either coefficient preserves the admissibility
condition. Therefore, the rational constants
$C_0=1.2148$ and $C_1=1.0955$ also satisfy
\eqref{eqn:ifr_admissible_constants}.

\subsection{Proof of Proposition~\ref{prop:ifr_fluid_gap}}

For each integer $B\geq2$, consider $B$ products consisting
of one product $h$ and $B-1$ additional products. Let
$\varepsilon_B=B^{-2}$ and
$a_B=\lfloor wB\rfloor/B$, so that $a_B\to w$.
The number of arriving customers is geometric on
$\mathbb{Z}_+$ with mean $B$, which is an IFR distribution.

Define the full-horizon demand means by
\[
\mu_h=B-(B-1)\varepsilon_B,
\qquad
\mu_i=\varepsilon_B
\quad(i\neq h).
\]
Each customer independently requests product $i$ with
probability $p_i=\mu_i/B$, and purchases it if it is
available. Otherwise, the customer leaves without purchasing.
Since $\sum_i p_i=1$, this defines a valid choice model with
\[
\pi(i,S)
=
p_i\mathbf{1}\{i\in S\}.
\]
It satisfies substitutability. Set the product revenues to
\[
r_h=1,
\qquad
r_i=\frac{a_B}{\varepsilon_B}
\quad(i\neq h).
\]

\paragraph{The fluid optimum.}
Offering the full assortment throughout the horizon generates
expected total demand exactly $B$, so assigning it fluid
weight one is feasible. Moreover, removing a product does not
increase demand for any other product. Therefore, the full
assortment maximizes unlimited-inventory expected revenue,
and
\begin{equation}
\label{eqn:ifr_gap_fluid_value}
\operatorname{OPT}_{\mathrm{FL}}
=
\mu_h+(B-1)a_B.
\end{equation}
Consequently,
\[
\frac{\operatorname{OPT}_{\mathrm{FL}}}{B}
\longrightarrow1+w.
\]

\paragraph{An upper bound on the optimal inventory revenue.}
Consider any feasible integral inventory vector, and let
$k\in\{0,\ldots,B\}$ denote the inventory of product $h$.
Bernoulli thinning of the geometric arrival count implies
that hypothetical demand for $h$ is geometric with mean
$\mu_h$. Its expected sales therefore equal
\[
\mu_h
\left[
1-\left(\frac{\mu_h}{1+\mu_h}\right)^k
\right].
\]
Every stocked product $i\neq h$ generates expected revenue
at most its unlimited-inventory expected revenue
$r_i\mu_i=a_B$.
Since at most $B-k$ such products can be stocked,
\[
\mathrm{Rev}(\boldsymbol{c})
\leq
\mu_h
\left[
1-\left(\frac{\mu_h}{1+\mu_h}\right)^k
\right]
+a_B(B-k).
\]

Writing $x=k/B$, the right-hand side divided by $B$
converges uniformly for $x\in[0,1]$ to
\[
\varphi(x)
=
1-e^{-x}+w(1-x).
\]
Indeed, $\mu_h/B\to1$ and
$B\log(\mu_h/(1+\mu_h))\to-1$.
Since $\varphi'(x)=e^{-x}-w$, the maximum is attained at
$x=-\log w=w$, with value
\[
\varphi(w)
=
1-w^2.
\]
It follows that
\[
\limsup_{B\to\infty}
\frac{\operatorname{OPT}_{\mathrm{SA}}}{B}
\leq
1-w^2.
\]

\paragraph{Matching lower bound.}
Now stock $k_B=\lfloor wB\rfloor$ units of product $h$
and one unit of each of $B-k_B$ additional products.
Each stocked product $i\neq h$ generates expected revenue
\[
r_i\Pr(G_{\varepsilon_B}\geq1)
=
\frac{a_B}{1+\varepsilon_B}.
\]
The resulting total expected revenue divided by $B$
converges to
\[
1-e^{-w}+w(1-w)
=
1-w^2,
\]
where we use $e^{-w}=w$.
Therefore,
\[
\frac{\operatorname{OPT}_{\mathrm{SA}}}{B}
\longrightarrow1-w^2.
\]
Combining this with~\eqref{eqn:ifr_gap_fluid_value},
we obtain
\[
\frac{\operatorname{OPT}_{\mathrm{SA}}}
{\operatorname{OPT}_{\mathrm{FL}}}
\longrightarrow
\frac{1-w^2}{1+w}
=
1-w,
\]
as required.

\section{Proofs from Section~\ref{sec:proof_dap}}

\subsection{Proof of Lemma~\ref{lem:dap_configuration_upper_bound}}

Fix an initial inventory vector
$\boldsymbol{c}\in\mathcal{C}_B$ and an admissible adaptive
personalized assortment policy $\sigma$.
Its initial product support
$I=\{i:c_i>0\}$ satisfies $|I|\leq B$, and every assortment
offered by the policy is a subset of $I$.

Let $K_t$ denote the type of customer $t$ and $S_t$ the
assortment offered to that customer. For every type $k$
with $\tau_k>0$ and assortment $S\subseteq I$, define
\[
y_{kS}
=
\frac{1}{\tau_k}
\mathbb{E}\left[
\sum_{t=1}^{T}
\mathbf{1}\{K_t=k,S_t=S\}
\right].
\]
Including the empty assortment, we have
$\sum_{S\subseteq I}y_{kS}=1$.
If $\tau_k=0$, set $y_{k,\emptyset}=1$.

Construct a distribution over configurations by independently
drawing $A_k$ according to $y_{kS}$ for each type $k$.
Thus, the configuration
$\boldsymbol{A}=(A_1,\ldots,A_K)$ has probability
$f_{\boldsymbol{A}}=\prod_{k=1}^K y_{k,A_k}$.
Every such configuration has
$I_{\boldsymbol{A}}\subseteq I$ and therefore belongs
to $\mathfrak{A}_B$.

Let $N_i$ denote the total sales of product $i$ under the
original inventory and adaptive policy. Since the current
customer's preference realization is independent of the
past, conditional on the current type and offered assortment,
\[
\mathbb{E}[N_i]
=
\sum_{k=1}^K\tau_k
\sum_{S\subseteq I}y_{kS}\pi^k(i,S).
\]
The type-specific marginals of the configuration distribution
therefore give
\[
\begin{aligned}
\sum_{\boldsymbol{A}\in\mathfrak{A}_B}
f_{\boldsymbol{A}}\mu_i^{\boldsymbol{A}}
&=
\sum_{k=1}^K\tau_k
\sum_{S\subseteq I}y_{kS}\pi^k(i,S)\\
&=
\mathbb{E}[N_i].
\end{aligned}
\]
Consequently,
\[
\sum_{\boldsymbol{A}}
f_{\boldsymbol{A}}D_{\boldsymbol{A}}
=
\sum_i\mathbb{E}[N_i]
\leq
\sum_i c_i
\leq B,
\]
and
\[
\sum_{\boldsymbol{A}}
f_{\boldsymbol{A}}V_{\boldsymbol{A}}
=
\sum_i r_i\mathbb{E}[N_i]
=
\mathrm{Rev}^{\mathrm{DAP}}(\boldsymbol{c},\sigma).
\]
Together with $\sum_{\boldsymbol{A}}f_{\boldsymbol{A}}=1$,
these relations establish feasibility and equality of
objective values. Taking the supremum over admissible
policies and maximizing over initial inventories proves
$U_{\mathrm{conf}}\geq\operatorname{OPT}_{\mathrm{DAP}}$.

\subsection{Proof of Lemma~\ref{lem:dap_approximate_fluid}}

For brevity, write $\alpha=\alpha_{\mathrm{JAC}}$ and
$U=U_{\mathrm{conf}}$.
The empty configuration is feasible, so $H(\lambda)\geq0$.
The fluid constraints imply
$U\leq\lambda B+H(\lambda)$ for every $\lambda\geq0$.
Hence, a configuration returned by the JAC oracle at
price $\lambda$ satisfies
\begin{equation}
\label{eqn:dap_oracle_lagrangian}
V_{\boldsymbol{A}}-\lambda D_{\boldsymbol{A}}
\geq
\alpha H(\lambda)
\geq
\alpha(U-\lambda B).
\end{equation}

First call the oracle at $\lambda=0$, obtaining a
configuration with revenue $V_0$ and demand $D_0$.
If $V_0=0$, then $H(0)=0$ and $U=0$, so the zero solution
suffices. If $D_0\leq B$, assigning weight one to this
configuration gives a feasible solution with value
$V_0\geq\alpha H(0)\geq\alpha U$.

It remains to consider $D_0>B$ and $V_0>0$.
Assigning weight $B/D_0$ to this configuration gives a
feasible fluid solution with positive value
$L=BV_0/D_0$. Thus, $0<L\leq U$.

Initialize a price interval with
$\lambda_-=0$ and $\lambda_+=r_{\max}$, where
$r_{\max}=\max_i r_i$.
Associate the first oracle configuration with the lower
endpoint and the empty configuration with the upper
endpoint. The latter is optimal at $r_{\max}$ because
all modified rewards are nonpositive.

At each iteration, query the oracle at the midpoint.
If the returned demand exceeds $B$, replace the lower
endpoint and its associated configuration. Otherwise,
replace the upper endpoint and its configuration.
The associated demands therefore always satisfy
$D_->B$ and $D_+\leq B$, and both endpoint configurations
satisfy~\eqref{eqn:dap_oracle_lagrangian}.
This invariant does not require monotonicity of the
oracle's choices.

Continue until
$\Delta=\lambda_+-\lambda_-\leq\varepsilon L/B$.
Set
\[
a=\frac{B-D_+}{D_--D_+}\in[0,1],
\]
and assign weights $a$ and $1-a$ to the configurations
associated with the lower and upper endpoints, respectively.
This mixture has total weight one and expected demand
$aD_-+(1-a)D_+=B$.

Applying~\eqref{eqn:dap_oracle_lagrangian} at both endpoints,
the mixture's objective value satisfies
\[
\begin{aligned}
F
&=
aV_-+(1-a)V_+\\
&\geq
\alpha U
+
a\lambda_-(D_--\alpha B)
+
(1-a)\lambda_+(D_+-\alpha B)\\
&=
\alpha U
+
\lambda_-(1-\alpha)B
+
(1-a)\Delta(D_+-\alpha B)\\
&\geq
\alpha U-\alpha B\Delta\\
&\geq
\alpha U-\alpha\varepsilon L\\
&\geq
\alpha(1-\varepsilon)U.
\end{aligned}
\]
The third line uses
$aD_-+(1-a)D_+=B$.
The fourth uses $\lambda_-\geq0$, $\alpha\leq1$,
and $D_+\geq0$.

The mixture is supported on at most two configurations.
Moreover, $L$ is a positive rational number with
polynomial-length encoding. Therefore, the required number
of bisection iterations is polynomial in the input size
and $\log(1/\varepsilon)$, completing the proof.

\subsection{Proof of Lemma~\ref{lem:dap_implementation}}

Fix a configuration $\boldsymbol{A}$ and an integral
inventory vector $\boldsymbol{c}$ supported on
$I_{\boldsymbol{A}}$.
Let $X_i^{\boldsymbol{A}}$ denote hypothetical demand for
product $i$ when every type-$k$ customer is offered the
full planned assortment $A_k$, without inventory restrictions.
We first establish
\begin{equation}
\label{eqn:dap_hypothetical_demand_bound}
\mathrm{Rev}^{\mathrm{DAP}}
(\boldsymbol{c},\sigma^{\boldsymbol{A}})
\geq
\sum_i r_i
\mathbb{E}[\min\{c_i,X_i^{\boldsymbol{A}}\}].
\end{equation}

\paragraph{The productwise coupling.}
Fix product $i$.
Let $K_t$ denote the customer type in period $t$, and
introduce independent uniform random variables
$U_1,U_2,\ldots$, independent of the arrival count and types.
Define
\[
p_{it}=\pi^{K_t}(i,A_{K_t}),
\qquad
Y_{it}=\mathbf{1}\{U_t\leq p_{it}\}.
\]
Then $\sum_{t=1}^{T}Y_{it}$ has the same distribution as
$X_i^{\boldsymbol{A}}$.

We couple the actual process to these variables as follows.
Given the history and type $K_t$, the actual assortment is
$S_t=A_{K_t}\cap\{j:c_{j,t}>0\}$.
Write $q_{it}=\pi^{K_t}(i,S_t)$.
Use $U_t$ to make the actual customer choose product $i$
exactly when $U_t\leq q_{it}$. On the complementary event,
use additional independent randomness to select among
the other outcomes according to their conditional choice
probabilities. This construction has the correct
conditional choice distribution.

Whenever product $i$ remains in stock, substitutability
implies $q_{it}\geq p_{it}$; if $i\notin A_{K_t}$,
both probabilities are zero. Therefore, a hypothetical
request for $i$ is accompanied by an actual sale of $i$
whenever inventory of $i$ remains available.
If $N_i(t)$ denotes cumulative actual sales through
customer $t$, induction gives
\[
N_i(t)
\geq
\min\left\{c_i,\sum_{s=1}^{t}Y_{is}\right\}.
\]
Evaluating at $T$ and taking expectations proves
$\mathbb{E}[N_i(T)]
\geq\mathbb{E}[\min\{c_i,X_i^{\boldsymbol{A}}\}]$.
The coupling is constructed separately for each product;
a simultaneous coupling across products is unnecessary.
Since revenues depend only on the purchased product,
summing these expected-sales comparisons proves
\eqref{eqn:dap_hypothetical_demand_bound}.

\paragraph{Deterministic and Poisson arrivals.}
When $T$ is deterministic, the variables $Y_{it}$ are
independent across periods, with
\[
\Pr(Y_{it}=1)
=
\sum_{k=1}^K\lambda_{kt}\pi^k(i,A_k).
\]
Thus, $X_i^{\boldsymbol{A}}$ is Poisson-binomial with mean
$\mu_i^{\boldsymbol{A}}$.
The convex-order comparison used in the proof of
Lemma~\ref{lem:revenue_lower_bound} therefore gives
\[
\mathbb{E}[\min\{c_i,X_i^{\boldsymbol{A}}\}]
\geq
g_{\mu_i^{\boldsymbol{A}}}(c_i).
\]

When $T$ is Poisson and types are independent and identically
distributed, each hypothetical customer selects product
$i$ with probability
$p_i=\sum_k\lambda_k\pi^k(i,A_k)$.
Poisson thinning gives
$X_i^{\boldsymbol{A}}\sim
\operatorname{Poisson}(\tau p_i)$, where
$\tau p_i=\mu_i^{\boldsymbol{A}}$.
The same comparison consequently holds with equality.

\paragraph{IFR arrivals.}
When $T$ is IFR and types are independent and identically
distributed, independently of $T$, we have
$X_i^{\boldsymbol{A}}\mid T
\sim\operatorname{Binomial}(T,p_i)$ with the same $p_i$.
The geometric convex-order argument in the proof of
Lemma~\ref{lem:ifr_revenue_lower_bound} gives
$X_i^{\boldsymbol{A}}
\leq_{\mathrm{cx}}G_{\mu_i^{\boldsymbol{A}}}$.
Consequently,
\[
\mathbb{E}[\min\{c_i,X_i^{\boldsymbol{A}}\}]
\geq
g_{\mu_i^{\boldsymbol{A}}}^{\mathrm{G}}(c_i).
\]
Substituting the appropriate demand comparison into
\eqref{eqn:dap_hypothetical_demand_bound} proves
the lemma in each arrival setting.

\end{appendices}

\end{document}